\documentclass[apj]{emulateapj}

\shorttitle{NaD excess objects}
\shortauthors{Jeong et al.}

\def\nad{Na\,D}
\def\fnad{\textit{fNaD}}
\def\mgb{Mg\,\textit{b}}
\def\fmgb{\textit{f}Mg\,\textit{b}}
\def\hb{H$\beta$}
\def\fhb{\textit{f}H$\beta$}
\def\fe{Fe\,5270}
\def\ffe{\textit{f}Fe\,5270}
\def\na1{Na\,{\small I} 8190}
\def\ebv{\textit{E(B\,$-$\,V)}}
\def\ebvs{\textit{E(B\,$-$\,V)$_{\rm diffuse}$}}
\def\ebvg{\textit{E(B\,$-$\,V)$_{\rm nebular}$}}

\begin{document}

\title{ON THE NATURE OF SODIUM EXCESS OBJECTS. I. DATA AND OBSERVED TRENDS }

\author{Hyunjin Jeong$^{1}$, Sukyoung K. Yi$^{2}$ Jaemann Kyeong$^{1}$, Marc Sarzi$^{3}$, Eon-Chang Sung$^{1}$, AND Kyuseok Oh$^{2}$ }
\affil{$^1$Korea Astronomy and Space Science Institute, Daejeon 305-348, Korea; hyunjin@kasi.re.kr\\
$^2$Department of Astronomy \& Yonsei University Observatory, Yonsei University, Seoul 120-749, Korea; yi@yonsei.ac.kr\\
$^3$Centre for Astrophysics Research, University of Hertfordshire, Hatfield Al10 9AB, UK}

\begin{abstract}
Several studies have reported the presence of sodium excess objects that have  
neutral atomic absorption lines at 5895\,\AA\ (\nad) and 8190\,\AA\  that are deeper 
than expected based on stellar population models that match the stellar continuum. 
Their origin is therefore hotly debated. van Dokkum \& Conroy proposed that 
low-mass stars (\,$\lesssim$\,0.3\,M$_{\odot}$) are more prevalent in massive 
early-type galaxies, which may lead to a strong \na1\ line strength. It is, however, 
necessary to test this prediction against other prominent line indices in optical 
wavelengths such as \nad, \mgb\ and \fe , which are measurable with a significantly 
higher signal-to-noise ratio than \na1. We newly identified roughly a thousand \nad\ 
excess objects (NEOs, $\sim$\,8\,\% of galaxies in the sample) based on the \nad\ 
line strength in the redshift range 0.00$\leqslant$\,$z$\,$\leqslant$\,0.08 from the 
Sloan Digital Sky Survey (SDSS) DR7 through detailed analysis of galaxy spectra, 
and then explored their properties. The novelty of this work is that galaxies were 
carefully identified through direct visual inspection of SDSS images, and we 
systematically compared the properties of NEOs and those of a control sample of 
normal galaxies in terms of \nad\ line strength. Note that the majority of galaxies with 
high velocity dispersion ($\sigma_{e}$\,$>$\,250\,km\,s$^{-1}$) show \nad\ excess. 
Most late-type NEOs have strong H$\beta$ line strengths and significant emission 
lines, which are indicative of the presence of young stellar populations. This implies 
that the presence of interstellar medium (ISM) and/or dust contributes to the increase 
in \nad\ line strengths observed for these galaxies, which is in good agreement with 
the earlier study of Chen et al. who used the \nad\ line index to study outflow activity 
in star-forming disk galaxies. In contrast, the majority of \textit{early-type} NEOs are 
predominantly luminous and massive systems, which is in agreement with the 
findings of van Dokkum \& Conroy. However, we find that models used to reproduce 
the \na1\ line strengths that adopt a bottom-heavy initial mass function (IMF) are 
\textit{not} able to reproduce the observed \nad\ line strengths. By comparing the 
observed \nad, \mgb\ and \fe\ line strengths with those of the models, we identify a 
plausible range of parameters that reproduce the observed values. In these models, 
the majority of early-type NEOs are ``$\alpha$-enhanced'' ([$\alpha$/Fe]$\sim$\,0.3), 
``metal-rich'' ([Z/H]\,$\sim$\,0.3) and especially ``Na-enhanced'' ([Na/Fe]\,$\sim$\,0.3).  
Enhanced Na abundance is a particularly compelling hypothesis for the increase 
in the strength of the \nad\ line index in our early-type NEOs that appear devoid of 
dust, both in their SDSS images and spectra.

\end{abstract}


\section{Introduction}
\label{sec:intro}

Deciphering galaxy spectra is important to understand the formation and evolution 
of galaxies. Prominent absorption features of galaxy spectra are widely used to 
study galaxy properties, because they reflect the average surface gravities, effective 
temperatures and metallicities of stellar components.

The behavior of spectral features in some galaxies in terms of sodium has received 
much attention as it has become known that some galaxies show enhanced \nad\ 
doublet strengths at 5890 and 5896\,\AA\  and \na1\ doublet strengths at 8183 and 
8195\,\AA.

\nad\ index is one of the strong absorption features of optical spectra, while \na1\ 
is one of the most prominent features in the near-IR (NIR) spectral range. Numerous 
studies have been performed over the last three decades to understand these lines.
It is now generally recognized that these lines are affected by the choice of an initial 
mass function (IMF), because these lines are strong in stars with a mass of less than 
0.3\,M$_{\odot}$. However, it is important to keep in mind that the \nad\ feature is 
more sensitive to Na-enhancement ([Na/Fe]) than gravity, and it can also be affected 
by the interstellar medium (ISM). Furthermore, both \nad\ and \na1\ absorption 
features are sensitive to age and metallicity.

What, then, causes \nad\ excess objects (hereafter NEOs)? The mechanism of \nad\ 
excess has been debated. One of the leading hypotheses is that \nad\ excess is 
related to the ISM, even though it is unclear what the dominant process is: 
galactic-scale gaseous outflows (galactic wind) in actively star-forming galaxies or 
active galactic nucleus (AGN) activity. It is well known that galactic winds are 
ubiquitous in star-forming galaxies \citep{hetal90}, and \citet{cetal10} argued that 
\nad\ absorption arises from cool gas in the disk, which is entrained in galactic wind. 
In contrast, \citet{letal11} investigated the properties of \nad\ line strengths in a sample 
of radio galaxies and concluded that AGN activity plays an important role in powering 
the outflow/heating phase of the feedback cycle. 

Another candidate is metal abundance. The discovery of non-solar abundance 
patterns in early-type galaxies was first made by \citet{o76} and \citet{p76}. They found 
extreme enhancement of \mgb\ and \nad\ features with respect to calcium and iron 
peaks, and concluded that this was a result of a higher metal abundance. Later, 
\citet{cvp86} measured the Na\,I, Ca\,II triplet, TiO and FeH features of 14 early-type 
galaxies and confirmed that these features are associated with metal abundance 
\citep[see also][]{ab89}. \citet{w98} also claimed that strong Na features are caused 
by an overabundance of [Na/Fe] (see also Worthey, Ingermann \& Serven 2011).

The latest compelling mechanistic theory is that of variation of the IMF. The stellar IMF 
is usually considered a universal function, but the possibility of a non-universal 
IMF has been raised by several authors 
\citep[see e.g.][]{d08,v08, tetal00,gfs09,tetal10,vc10}. 
One of the first observational attempts to constrain the low-mass portion of the IMF 
was  made by \citet{st71}. They compared the observed line strengths of some IMF 
sensitive lines such as  Na\,I and TiO in the centers of M31, M32 and M81 with the 
population synthesis models and claimed a substantial fractional contribution by dwarf 
stars to the integrated light of galaxies. \citet{c78} also measured the strengths of \na1\ 
and TiO in the centers of M31 and M32 and concluded  that the IMF is equal to that of 
the Galactic disk. However, \citet{ff80} argued for a bottom-heavy (dwarf-rich) IMF, 
despite the fact that their results were also based on the \na1\ line strengths in M31 
and M32. Indeed, it is a challenge to estimate the number of low-mass stars from an 
integrated spectrum, because low-mass stars are too faint to have a strong influence 
on the integrated spectrum.

The debate over IMF variation, however, was reopened by \citet{setal02}. They found 
an anti-correlation between the strength of the Ca\,II triplet region at  8500\,\AA\ and 
velocity dispersion for elliptical galaxies and concluded that bottom-heavy IMFs are 
favored \citep[see also][]{cetal03}. Recently, \citet{vc10} reported observing the 
near-infrared \na1\ doublet in the spectra of massive early-type galaxies, and claimed 
that this excess can be explained by the bottom-heavy IMF \citep[see also][]{vc12}. If 
correct, massive early-type galaxies should possess relatively more low-mass stars, 
and this would affect the mass-to-light ratio ($M/L$) of galaxies. \citet{mcetal12} claimed 
that there was systematic variation in the IMF in early-type galaxies as a function of 
stellar \textit{M/L} ratios based on detailed dynamical modeling of ATLAS$^{3D}$ 
early-type galaxies \citep[see e.g.][]{mcetal11}, but the expected maximum slope of the 
IMF is about 2.8 (x\,$\sim$\,$-$2.8), which is lower than that proposed by \citet{vc10}, 
and there are many high velocity dispersion galaxies that prefer a Salpeter IMF 
\citep{s55}.   

Three origins of \nad\ excess have been hypothesized: (1) the ISM, (2) metallicity 
and (3) a bottom-heavy stellar IMF. To determine which of the above is correct, we 
explore the properties of NEOs in the redshift range 
0.00\,$\leqslant$\,$z$\,$\leqslant$\,0.08 from the seventh data release of the Sloan 
Digital Sky Survey \citep[SDSS;][]{aetal09}, with spectral measurements retrieved from 
\citet{ossy11}. Almost all previous studies have focused on either early-type or late-type 
galaxies. However, it is important to investigate \nad\ excess according to galaxy 
morphology. The novelty of this paper is that we simultaneously inspect images to 
identify the morphology of galaxies and explore the photometric and spectroscopic 
properties of sample galaxies based on homogeneous data sets. Therefore, the 
catalogue presented here is a useful reference sample of NEOs with morphological 
information.

In this paper, we focus on the \nad\ feature of galaxies. In Section~\ref{sec:data}, we 
briefly summarize the photometric and spectroscopic data and describe the sample and 
our method for selecting NEOs. The properties of early-type and late-type NEOs are 
presented in Sections~\ref{sec:results_ETG} and \ref{sec:results_LTG}, respectively. 
Finally,  we discuss the origin of NEOs in Section~\ref{sec:discussion}.

\begin{table}[t]
 \begin{center}
 \caption{Summary of sampling criteria}
  \begin{tabular}{l l}
  \hline\hline
    \multicolumn{1}{l}{Criterion} & Explanation\\
  \hline
     0.00 $\leqslant$ $z$ $\leqslant$ 0.08  & Limit redshift for morphological classification\\
     $M_r$ $<$ $-$20.5   	& The absolute r-band magnitude cut for\\
     					& a volume-limited sample\\ 
     S/N $>$ 20                     & Guarantee the quality of spectroscopic data\\
     rN/sN$^a$ $\leqslant$ 1.5      & Guarantee the quality of the fit\\
                                               & to the stellar continuum\\
  \hline
  \end{tabular}
  \label{tab:sample_crit}
  \end{center}
  \bf{~~Notes.}
  \tablenotetext{a}{\citealt{ossy11}}
  \end{table}

\section[]{Data Construction}
\label{sec:data}

\subsection{Sloan Digital Sky Survey (SDSS)\\ and OSSY Catalogue}
\label{sec:sdss}

SDSS has established the largest and most homogeneous database of both photometric 
($u$, $g$, $r$, $i$ and $z$ bands) and spectroscopic data for about one million galaxies.  
However, it is known that the SDSS pipeline spectroscopic data measurements still have 
a few crucial defects. The most serious problem is that the pipeline values for the 
absorption line strength are contaminated with nebular emissions. 
\citet[][hereafter OSSY]{ossy11} recently released new and improved absorption and 
emission-line measurements for the SDSS galaxies using the Gas AND Absorption Line 
Fitting routine (\texttt{GANDALF};  \citealt{setal06}) and the penalized pixel-fitting code 
(\texttt{pPXF};  \citealt{ce04}). Furthermore, they provided information about the best-fit 
models for the SDSS spectra. We start the analysis based on this OSSY catalogue.
 
We obtain photometric values from the Catalog Archive Server for DR7. Note that we use 
Petrosian and model magnitudes to estimate galaxy luminosities and colors, respectively. 
We also deredden the colors with respect to Galactic extinction using dust maps provided 
by the SDSS pipeline (Schlegel et al. 1998) and apply a \textit{k}-correction based on 
simple two-population (young plus old) modeling of the observed optical colors. Details 
of the \textit{k}-correction are described more fully in Section 3 of \citet{yetal12}. 

\subsection{Sample Selection}
\label{sec:sample}

We begin by selecting all sample galaxies with spectra in the redshift range 
0.00\,$\leqslant$\,$z$\,$\leqslant$\,0.08 and apply the absolute r-band magnitude cut-off 
of $-$20.5 to obtain a volume-limited sample. Furthermore, only galaxies with a 
signal-to-noise (S/N) ratio above 20 and a quality assessment value based on continuum 
fit (rN/sN, see section 3 of OSSY) below 1.5 are included in our sample to guarantee high 
quality spectroscopic data and model fit to the stellar continuum. The total number of 
galaxies in our sample is 20\,571. Our selection criteria are summarized in 
Table~\ref{tab:sample_crit}.

 
\subsection{Measurement of \nad\ Excess}
\label{sec:fNaD}

To find NEOs, we define a new index, \fnad, which quantifies the \nad\ excess as follows:

\begin{eqnarray}
fNaD = \frac{\rm{Na\,D\,(Observed) - Na\,D\,(Model)}} {\rm{Na\,D\,(Model)}},
\end{eqnarray}
where \nad\,(Observed) is the observed \nad\ line strength and \nad\,(Model) is the 
expected model \nad\ line strength. For example, a \fnad\ value equal to 1 indicates that 
the observed absorption strength of \nad\ is two times stronger than that expected for the 
best-fit model. Details of the continuum fitting and absorption line measurements are 
described in OSSY, so we only briefly summarize the process here. The OSSY team first 
separated the contributions of the stellar continuum and ionized-gas emission to the 
observed spectrum by matching them only for emission-free regions using a set of stellar 
templates. They then measured the emission-line strengths by fitting the stellar templates 
and Gaussian emission-line templates to the data. During this process, they used both the 
stellar population synthesis models of \citet{bc03} and the MILES stellar templates 
\citep{sbetal06}, and calculated internal reddening due to dust (see 
Section~\ref{sec:ebv_ETG} for details). Finally, they subtracted the emission-line spectrum 
from the observed one to get the clean absorption line spectrum, and then measured the 
absorption line indices from this cleaned spectrum following standard line-strength index 
definitions.

\begin{figure}
\begin{center}
\includegraphics[width=0.45\textwidth]{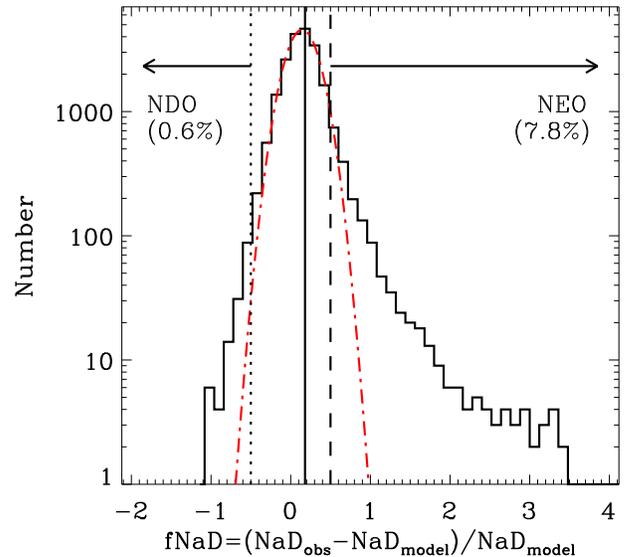}
\caption{Distribution of \fnad. The peak value is shown by the black solid line. The red 
dot-dashed line is a Gaussian fit to the data. From the total sample of 20\,571 galaxies, 
we selected a subsample of 1\,603 galaxies with significant \nad\ excess 
(\fnad \,$\geqslant$\,0.5, dashed line). Note that \nad\ deficient objects (NDO) also exist 
(\fnad \,$\leqslant$\,$-$0.5, dotted line). A color version of this figure is available in the 
online journal}.
\label{fig:fNaD}
\end{center}
\end{figure}

In Figure~\ref{fig:fNaD}, we plot the distribution of \fnad\ for a total of 20\,571galaxies. 
Interestingly, the distribution is skewed with a tail toward higher values. We fit a Gaussian 
curve to the \nad\ excess index distribution and plot it using the red dot-dashed line to 
dramatize the presence of NEOs. Moreover, the peak (vertical solid line) in the distribution 
does not correspond to zero. In other words, more galaxies tend to have positive \fnad\ 
values.
 
We adopt a conservative limit of \fnad \,$=$\,0.5 to pick out galaxies showing \nad\ excess. 
This demarcation for identifying NEOs (\fnad\ \,$\geqslant$\,0.5) is indicated by the vertical 
dashed line in Figure~\ref{fig:fNaD}. The overall fraction of NEOs within our criterion is 
roughly 7.8\,\% (1\,603/20\,571). Interestingly, \nad\ deficient objects (NDO) also exist. If 
we adopt a limit of \fnad \,$=$\,$-$0.5 (vertical dotted line) to select NDOs 
(\fnad \,$\leqslant$\,$-$0.5), the overall fraction of NDOs is  about 0.6\,\% (120/20\,571). 
We note that most NDOs are strong emission-line galaxies. \citet{ba86} indeed 
reported that a few late-type galaxies have a \nad\ emission line. This implies that the \nad\ 
absorption feature can be affected by the \nad\ emission line, and that continuum fits may 
be influenced by emission lines.

An important objective of this paper is a systematic comparison of the properties of NEOs 
and those of a control sample of galaxies that do not show \nad\ excess. We create a 
control sample of $\sim$\,1\,600 galaxies in the \nad\ excess range 
0.0\,$\leqslant$\,\fnad \,$\leqslant$\,0.1, based on the pick value of \fnad\ of about 0.1 (see 
Figure~\ref{fig:fNaD}). The classification scheme is summarized below: 
\begin{equation}
  \begin{array}{ll} \fnad \,\geqslant 0.5  &  \quad \rm{\nad\ excess~objects~(NEOs)}\\
                               0.0\,\leqslant\,\fnad \,\leqslant\,0.1    & \quad \rm{Control~sample}\\
                               \fnad \,\leqslant -\,0.5   & \quad \rm{\nad\ deficient~objects~(NDOs)}
  \end{array} 
\end{equation}

\begin{figure*}[t]
\begin{center}
\includegraphics[width=0.85\textwidth]{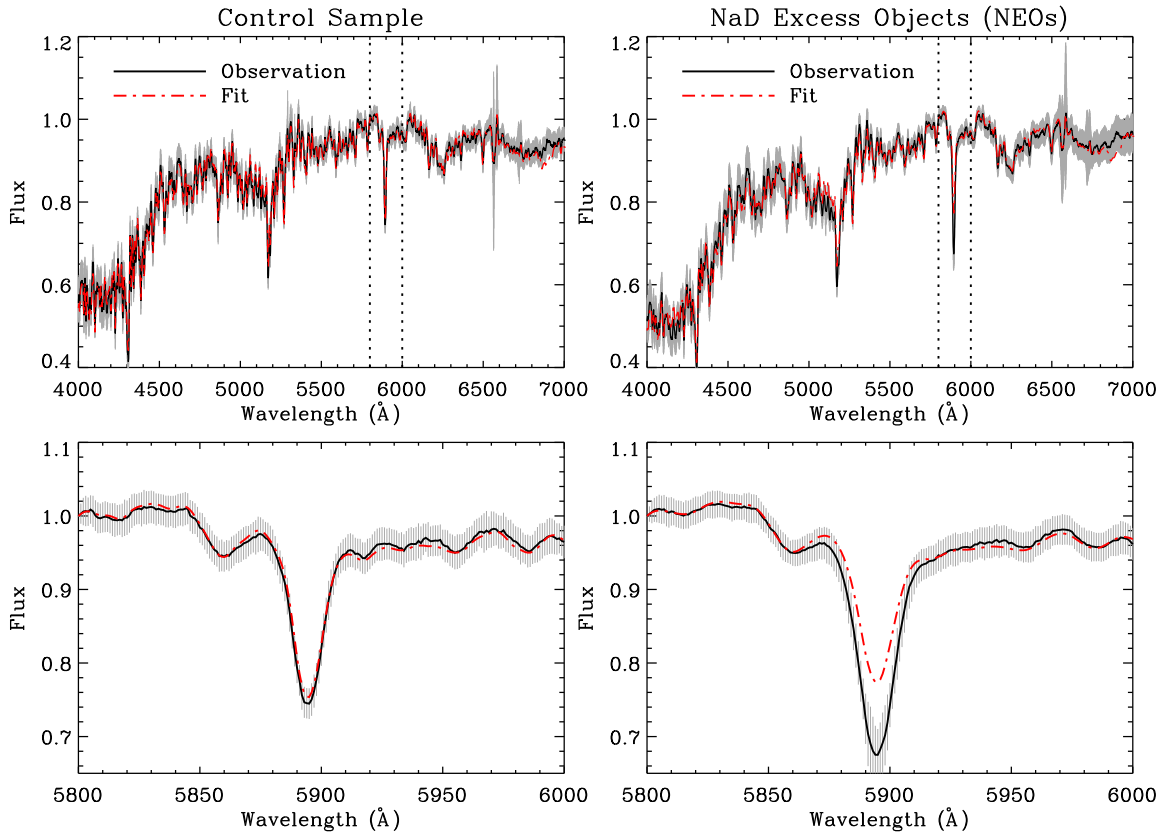}
\caption{Sample SDSS stacked spectra of our control sample galaxies and \nad\ excess 
objects. The observed spectra, including the 1\,$\sigma$ scatter, are shown by the black 
solid lines, together with their best fits (red dot-dashed lines). The bottom figures show 
the spectra in the regions of the \nad\ feature. The spectra were normalized at 5800\,\AA.
A color version of this figure is available in the online journal.}
\label{fig:sample_sed}
\end{center}
\end{figure*}

\begin{table}[t]
 \begin{center}
 \caption{Results of Morphological Classification}
  \begin{tabular}{llr}
  \hline\hline
  \multicolumn{1}{l}{Classification} & \multicolumn{1}{l}{} &  \multicolumn{1}{r}{Number} \\
  \hline
     Control Sample (1036) & & \\
                                                & Early-type galaxies & 694\\
                                                & Late-type galaxies & 342\\
     \hline
     NEOs (963) & & \\                                
                           & oETG$^a$ & 206 \\
                           & pETG$^b$ & 347 \\
                           & oLTG$^c$ & 56 \\
                           & pLTG$^d$ & 354 \\     
  \hline
  \end{tabular}
  \label{tab:sample_number}
  \end{center}
  \bf{~~Notes.}
  \tablenotetext{a}{ordinary early-type \nad\ excess objects}
  \tablenotetext{b}{peculiar early-type \nad\ excess objects}
  \tablenotetext{c}{ordinary late-type \nad\ excess objects}
  \tablenotetext{d}{peculiar late-type \nad\ excess objects}
 \end{table}

Observed stacked spectra (black solid lines) of sample galaxies with their stacked best fits 
(red dot-dashed lines), normalized at 5800\,\AA, are shown in Figure~\ref{fig:sample_sed}. 
We construct the stacked galaxy spectra by selecting $\sim$\,300 early-type looking galaxies 
in the \fnad\ range 0.0\,$\leqslant$\,\fnad \,$\leqslant$\,0.1 (for the control sample) and 
\fnad \,$\geqslant$\,0.5 (for the NEOs), respectively, and we also estimate the 1\,$\sigma$ 
scatter (grey lines). The overall observed spectra match the models well. However, there is 
a marked discrepancy in fit in the specific index region for sodium near 5900\,\AA. The vast 
majority of the spectra are well matched (left), but some galaxies (NEOs, \,$\sim$\,7.8\,\%) 
show a much stronger observed line index than expected based on the best-fit model (right). 
We note that the expected model \nad\ line strengths of the control sample and NEOs are 
almost the same.  
                  
\subsection{Visual Classification}
\label{sec:visual}
 
As the final step, we perform visual classification of the sample. Images used for 
morphological classification are those provided by the SDSS DR7 release. The sample 
galaxies showing \nad\ excess are carefully assigned to four main classes: (1) ordinary 
early-type galaxies (oETGs), (2) peculiar early-type galaxies (pETGs), (3) ordinary late-type 
galaxies (oLTGs) and (4) peculiar late-type galaxies (pLTGs). Note that galaxies that exhibit 
any asymmetric features (e.g. shells and tails), dust patches or dust lanes (especially for 
early-type galaxies)  are classified as peculiar galaxies, that is, categories (2) or (4).  Our 
classification scheme, therefore, is to construct robust ordinary samples (categories (1) and 
(3)). There are, however, little differences among late-type galaxies. We use only two 
categories for the control sample: early-type galaxies and late-type galaxies.  




\begin{figure}[t]
\begin{center}
\includegraphics[width=0.48\textwidth]{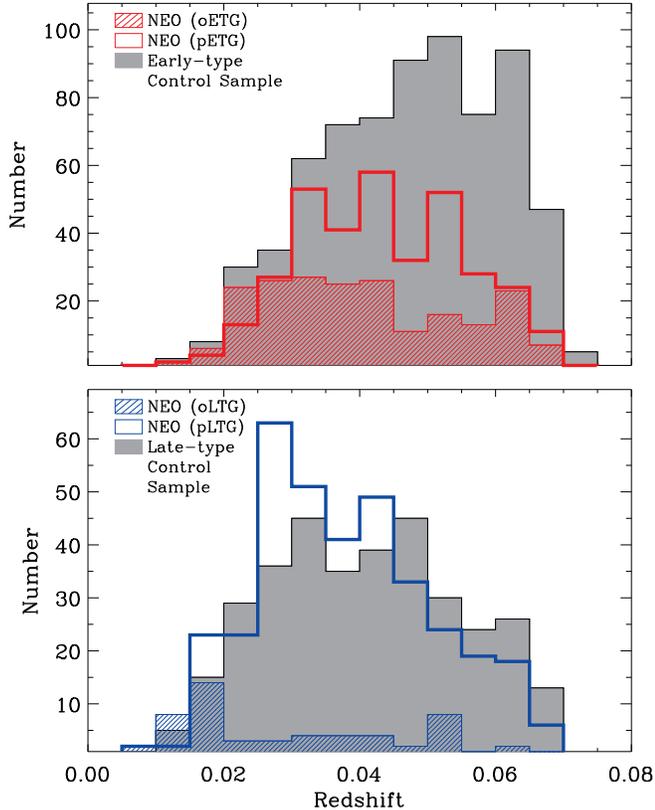}
\caption{\textit{Top}: Distribution of redshift for our visually classified ordinary (red 
hashed) and peculiar (red solid) early-type NEOs. The grey filled histogram 
represents the early-type control sample. \textit{Bottom}: Distribution of redshift for 
our visually classified ordinary (blue hashed) and peculiar (blue solid) late-type 
NEOs. The grey filled histogram represents the late-type control sample.
A color version of this figure is available in the online journal.}
\label{fig:redshift}
\end{center}
\end{figure}

Additionally, we keep the galaxy as a sample if we can precisely morphologically classify 
it. When we are unsure of the classification, the galaxy is classified as an unknown galaxy. 
This process yields a final sample of 206 reliably classified oETGs, 347 pETGs, 56 oLTGs, 
354 pLTGs, 694 early-type control samples and 342 late-type control samples. Details are 
listed in Table~\ref{tab:sample_number}. 
Figure~\ref{fig:redshift} shows a plot of the redshift distributions of NEOs and control 
sample galaxies. Note that we initially construct a volume-limited sample, but using an S/N 
cut and visual classification leads to exclusion of  a fair fraction of more distant galaxies 
from our sample. It is also interesting to note that only 15\,\% of galaxies that show extreme 
\nad\ excess (\fnad\,$\geqslant$\,1.0) are classified as early-type galaxies.

\section[]{Properties of early-type\\ \nad\ excess objects}
\label{sec:results_ETG}

\subsection{Color-Magnitude Relation}
\label{sec:cmr_ETG}

The color-magnitude relation (CMR) is widely used to study the star formation history of 
galaxies. Galaxies are populated in three main regions of the color-magnitude diagram: 
the red sequence, the blue cloud and the green valley in between. It is known that the 
optical CMR of early-type galaxies shows a small scatter around the mean relation. In 
other words, almost all of the early-type galaxies are on the red sequence. 

\begin{figure}[t]
\begin{center}
\includegraphics[width=0.48\textwidth]{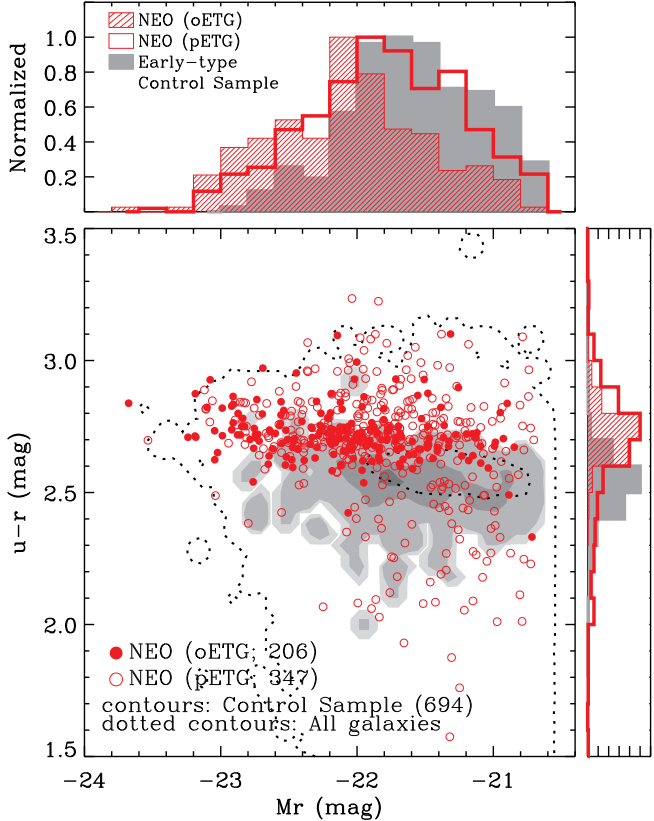}
\caption{\textit{Top}: Distribution of $r$-band absolute magnitudes for our early-type 
sample. The grey filled histogram represents the early-type control sample, while the 
red hashed and solid histograms indicate the ordinary (oETGs) and peculiar (pETGs) 
early-type \nad\ excess objects. \textit{Bottom-left}: Color-magnitude relation for our 
early-type \nad\ excess objects. The ordinary  (oETGs) and peculiar (pETGs) early-type 
NEOs are plotted as red filled and open circles,  respectively. The contours of the 
early-type control sample (grey shaded) and all sample galaxies (dotted)  are shown 
for comparison.
\textit{Bottom-right}: Distribution of u$-$r colors for our early-type sample.
A color version of this figure is available in the online journal.}
\label{fig:cmr_ETG}
\end{center}
\end{figure}

Figure~\ref{fig:cmr_ETG} shows the u$-$r CMR. The u$-$r CMR is a particularly good tool 
for tracking the presence of young stellar populations. Grey contours, red filled circles and 
red open circles indicate the early-type control sample, ordinary  and peculiar early-type 
NEOs, respectively. A cursory glance at this diagram shows that most ordinary early-type 
NEOs (oETGs, filled circles) reside in the well-defined red sequence, while peculiar 
early-type NEOs (pETGs, open circles) have a wide color baseline. This is because our 
morphological classification scheme for peculiar galaxies allows for a wide variety of 
galaxies, ranging from dusty red galaxies to star-forming blue galaxies possessing shells, 
tidal features and signatures of recent star formation. The M$_r$ and u$-$r color 
distributions of the ordinary (red hashed) and peculiar (red solid) early-type NEOs 
compared to those of their control-sample counterparts (grey filled) are shown in 
Figure~\ref{fig:cmr_ETG}. Ordinary early-type NEOs are more luminous and optically 
redder than the control sample. Meanwhile, the majority of peculiar early-type NEOs are 
also slightly more luminous and redder than the control sample, but vary in luminosity and 
color.

\subsection{Velocity Dispersion and Stellar Mass}
\label{sec:vd_ETG}

Stellar velocity dispersion is one of the important physical parameters of galaxies, and can 
be used to estimate galaxy mass by applying the virial theorem or a related method. 
Because SDSS fibers have a fixed diameter, we apply aperture correction to the observed 
velocity dispersion using the following formula of \citet{cetal06}:

\begin{eqnarray}
\sigma_{\rm e} = (\,\frac{R_{\rm fiber}}{R_{\rm e}}\,)^{0.066\,\pm\,0.035} \;\sigma_{\rm fiber},
\end{eqnarray}
where $R_{\rm fiber}$ is the aperture radius of the SDSS fiber (1.5\,$''$) and 
$R_{\rm e}$ is the effective radius.

The stellar mass of each galaxy is also derived from its color and luminosity using the 
following formula of \citet{betal03}:
\begin{eqnarray}
log\left(\frac{M_*}{M_\odot}\right)=-0.306+1.097(g-r)-0.4(M_{\rm r,*}-M_{\rm r,{\odot}}).
\end{eqnarray}

\begin{figure}[t]
\begin{center}
\includegraphics[width=0.49\textwidth]{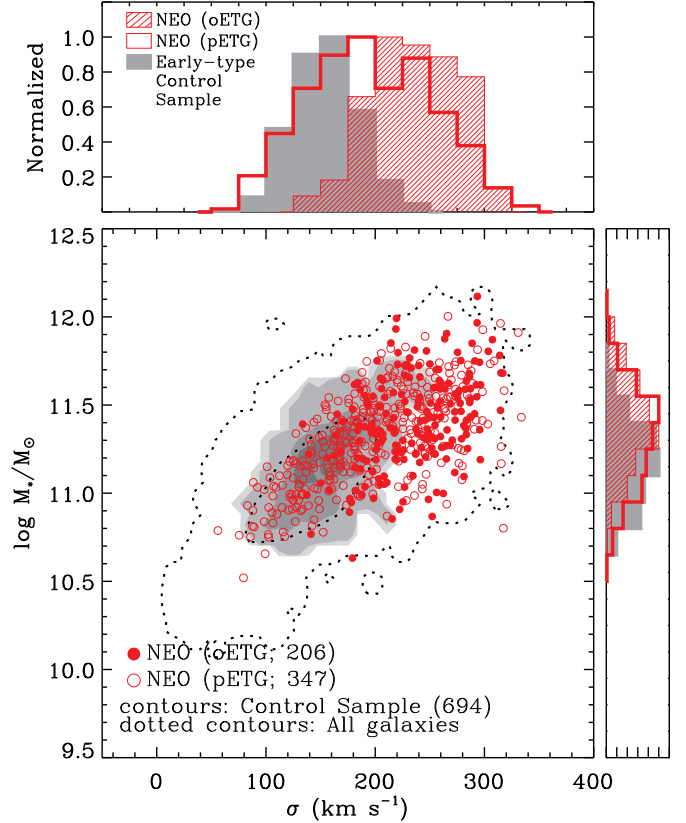}
\caption{\textit{Top}: Distribution of velocity dispersions for our early-type sample. 
Early-type control sample, ordinary and peculiar early-type NEOs are represented 
as grey filled, red hashed and red solid histograms, respectively.
\textit{Bottom-left}: Stellar mass of our early-type \nad\ excess objects as a function 
of velocity dispersion. The ordinarily early-type NEOs are shown using red filled 
circles, while peculiar early-type NEOs are plotted as red open circles. The contours 
of the early-type control sample (grey shaded) and all sample galaxies (dotted) 
are shown for comparison.
\textit{Bottom-right}: Distribution of stellar masses for our early-type sample.
A color version of this figure is available in the online journal.}
\label{fig:vd_ETG}
\end{center}
\end{figure}

Figure~\ref{fig:vd_ETG} shows the stellar mass of our early-type sample as a function of 
velocity dispersion. Both ordinary (red filled circles) and peculiar (red open circles) 
early-type NEOs generally tend to have higher velocity dispersions and correspondingly 
bigger masses than the majority of the control sample (grey contours). We also plot the 
distributions of velocity dispersion and stellar mass for ordinary (red hashed) and peculiar 
(red solid) early-type NEOs. The histograms for the control sample (grey filled) are also 
provided for comparison. We find again that ordinary early-type NEOs have significantly 
higher velocity dispersions and are relatively more massive systems than the average 
early-type control sample. Furthermore, we checked the velocity dispersions of all sample 
galaxies in the \nad\ excess range 0.0\,$\leqslant$\,\fnad\,$\leqslant$\,0.1 and find that 
only 0.5\,\% (17/3\,461) of galaxies without a significant \nad\ excess have velocity 
dispersions greater than 250 km\,s$^{-1}$. This implies that the majority of high velocity 
dispersion galaxies show \nad\ excess. Our results are consistent with those of 
\citet{setal12} and \citet{fetal13}, who found a correlation between \na1\ line strengths and 
velocity dispersions. Meanwhile, peculiar early-type NEOs have a wide range of velocity 
dispersions.


\subsection{Emission Line Diagnostics: \\ Star Formation and AGN Activity}
\label{sec:bpt_ETG}

The ratios of emission lines can be used to distinguish various classes of emission line 
galaxies and ionization mechanisms. One of the most frequently used methods is the 
BPT diagram, proposed by \citet{bpt81}, on the basis of 
[O\,{\small III}]\,$\lambda$\,5007/H$\beta$\,$\lambda$\,4861 and 
[N\,{\small II}]\,$\lambda$\,6583/H$\alpha$\,$\lambda$\,6563 ratios. This method allows 
the classification of galaxies into star-formation-dominant and AGN-dominant galaxies.

\begin{figure*}[t]
\begin{center}
\includegraphics[width=0.95\textwidth]{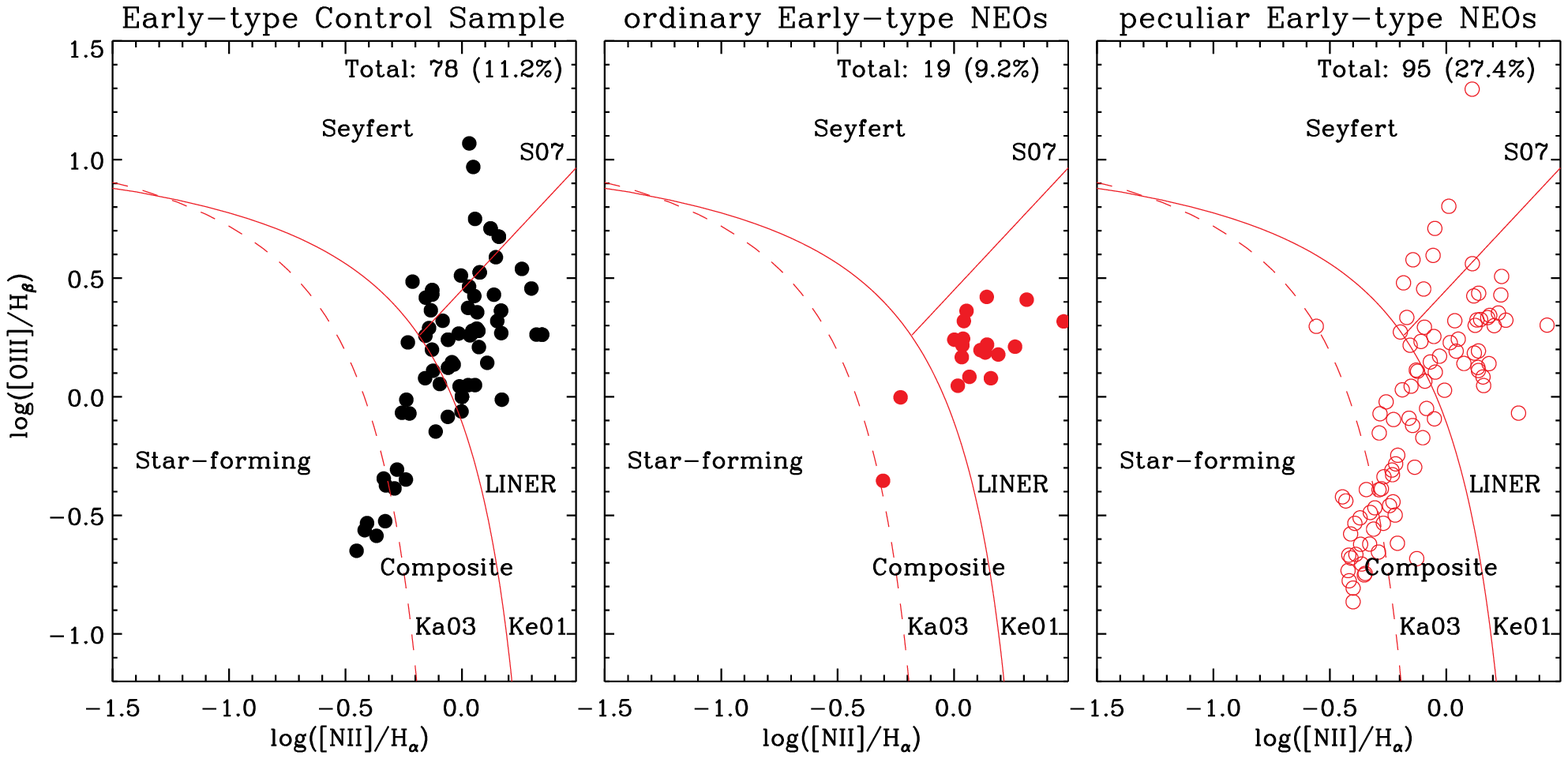}
\caption{ Emission-line ratio diagram for our early-type sample. Only galaxies with 
an A/N greater than 3 for all four lines are shown. The curve labeled Ka03 is the 
empirical purely star-forming limit of \citet{ketal03}, while the curve labeled Ka01 is 
the theoretical maximum starburst model of \citet{ketal01}. The straight line labeled 
S07 from \citet{setal07} divides Seyferts and LINER AGNs.Black filled circles indicate 
the early-type control sample (left). Ordinarily early-type NEOs are shown using red 
filled circles (middle), while peculiar early-type NEOs are plotted as red open circles 
(right). A color version of this figure is available in the online journal.}
\label{fig:bpt_ETG}
\end{center}
\end{figure*}

BPT diagrams for the early-type control sample (left, black filled circles), ordinary early-type 
NEOs (middle, red filled circles) and peculiar early-type NEOs (right, red open circles) are 
shown in Figure~\ref{fig:bpt_ETG}.  We note that galaxies in which all four emission lines 
are detected with an amplitude-over-noise (A/N from \texttt{GANDALF}, which is similar to 
the signal-to-noise ratio) greater than 3 are classified as either `star-forming', `composite' 
(i.e. hosting both star formation and AGN activity), `Seyfert' or `LINER' using the demarcation 
lines of \citet[dashed curve]{ketal03}, \citet[solid curve]{ketal01,ketal06} and 
\citet[straight line]{setal07}. Galaxies without a detection in all four lines are classified as 
`quiescent' galaxies. Details are provided in Table~\ref{tab:bpt_ETG}. 

Early-type NEOs exhibit a two-fold higher fraction of strong emission lines ($\sim$\,21$\,\%$) 
than the early-type control sample ($\sim$\,11$\,\%$). Moreover, the fraction of early-type 
NEOs classified as `star-forming' is four-fold that of the control counterpart. The reason why 
early-type NEOs show a higher fraction of strong emission lines is mainly because of 
peculiar early-type NEOs. Most ordinary early-type NEOs are virtually free of emission lines 
($\sim$\,91\,\%). It is also worth noting that the overwhelming bulk of emission in ordinary 
early-type NEOs is `LINER' AGN, which implies that ordinary early-type NEOs do not host 
both active star formation and powerful AGNs. 

\begin{table*}[t]
 \begin{center}
  \caption{Results of Spectral Line Classification for Early-type Galaxies}
  \begin{tabular}{llrrrr}
  \hline \hline
 \multicolumn{1}{c}{Classification} & & early-type control sample & early-type NEO & oETGs$^a$ & pETGs$^b$ \\
 \hline
 Quiescent            &           &   88.8\% (616) &  79.4\% (439) & 90.8\% (187) & 72.6\% (252) \\
 Strong emission &           & 11.2\% (78)     &  20.6\% (114) &   9.2\% (19)    & 27.4\% (95) \\
       & Star-forming            &  1.0\% (7)         &    4.0\% (22)   &   0.0\% (0)      &   6.3\% (22) \\
       & Composite               &  2.0\% (14)      &     5.8\% (32)   &   1.0\% (2)      &   8.7\% (30) \\
       &  Seyfert                     &  2.2\% (15)      &     1.4\% (8)     &   0.0\% (0)      &   2.3\% (8) \\
       &  LINER                      &  6.0\% (42)      &     9.4\% (52)   &   8.2\% (17)    & 10.1\% (35) \\ 
 \hline
 \end{tabular}
 \label{tab:bpt_ETG}
 \end{center}
 \bf{~~Notes.}
 \tablenotetext{a}{ordinary early-type \nad\ excess objects}
 \tablenotetext{b}{peculiar early-type  \nad\ excess objects}
\end{table*}

\begin{figure*}[t]
\begin{center}
\includegraphics[width=0.85\textwidth]{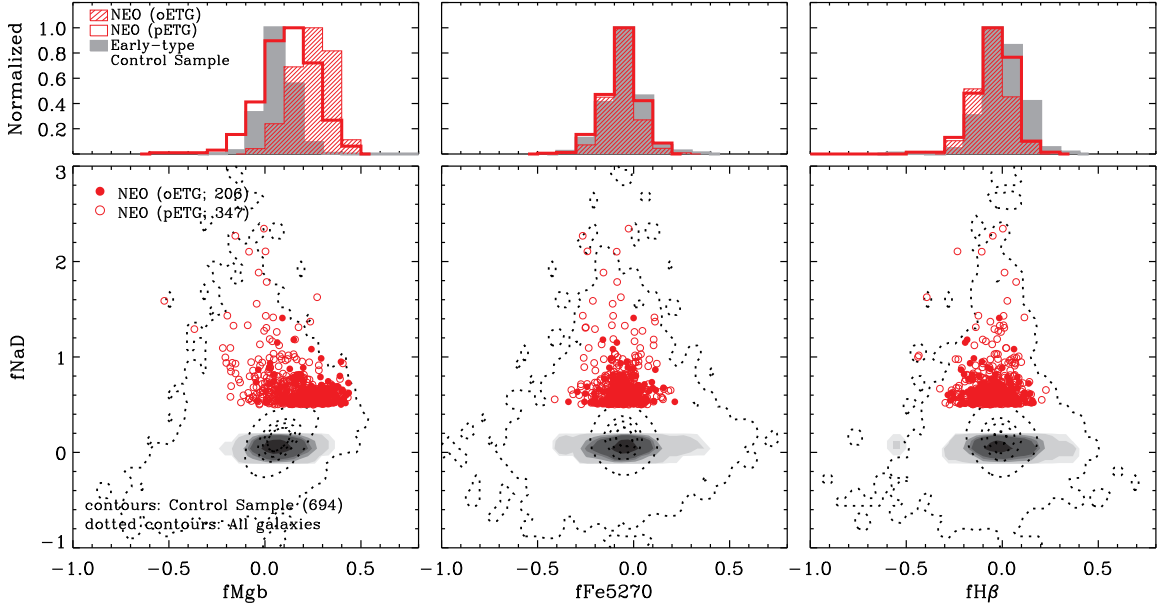}
\caption{\textit{Top}: \mgb, \fe\ and \hb\ excess index histograms for our early-type 
control sample (grey filled), ordinary (red hashed) and peculiar (red solid) early-type 
\nad\ excess objects. \textit{Bottom}: \fnad$-$excess index plots for the early-type 
control sample (grey shaded contours), ordinary (red filled circles) and peculiar (red 
open circles) early-type NEOs. For comparison, the contours of all sample galaxies 
are shown as dotted lines.
A color version of this figure is available in the online journal.}
\label{fig:fNaD_ETG}
\end{center}
\end{figure*}

\begin{figure*}[t]
\begin{center}
\includegraphics[width=0.85\textwidth]{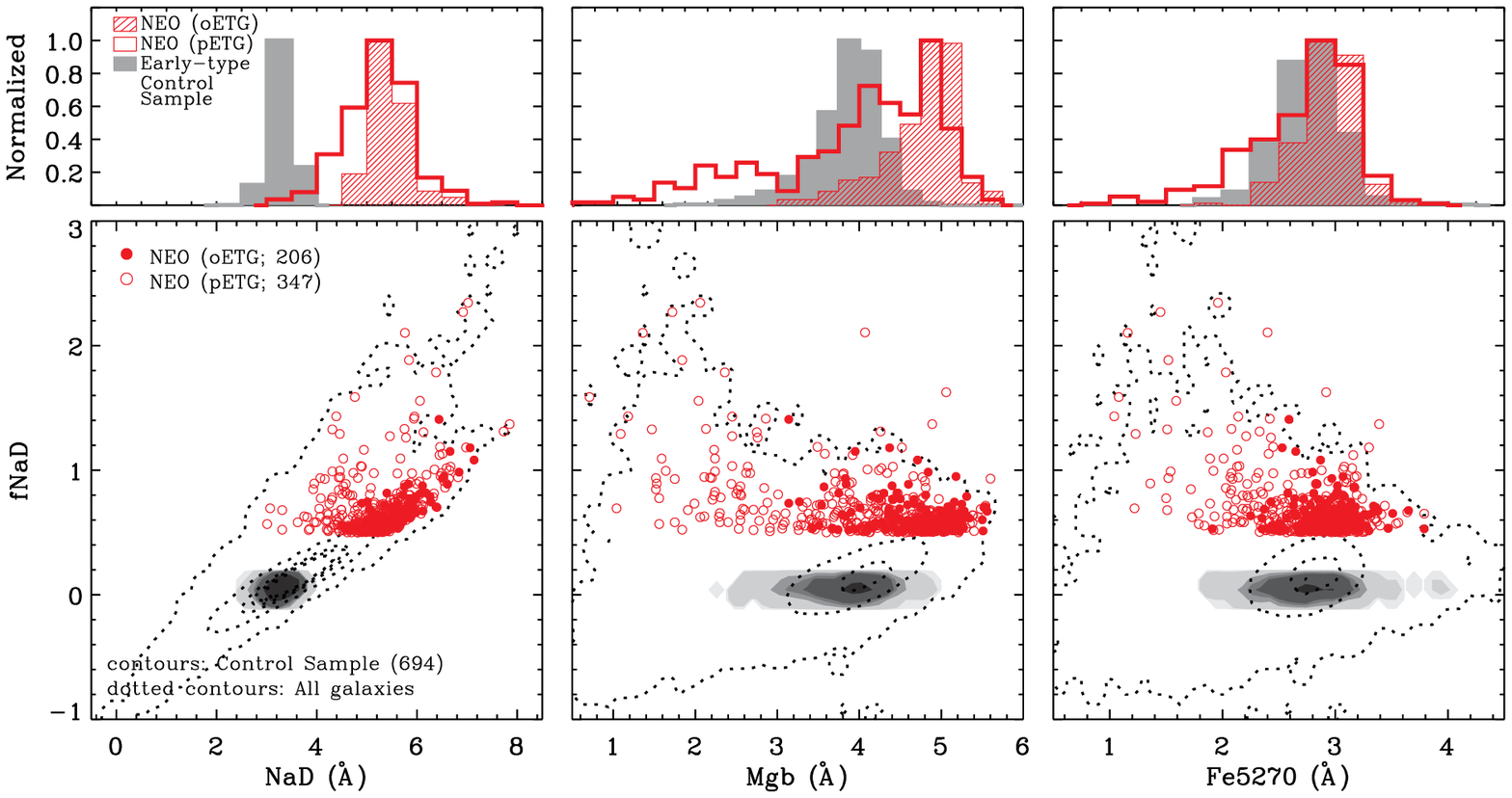}
\caption{\textit{Top}: Histograms of the \nad, \mgb\ and \fe\ line strengths for the 
early-type control sample (grey filled), ordinary (red hashed) and peculiar (red solid) 
early-type \nad\ excess objects. \textit{Bottom}: \fnad$-$line strength index plots for 
the early-type control sample (grey shaded contours), ordinary (red filled circles) and 
peculiar (red open circles) early-type NEOs. The contours of all sample galaxies are 
also shown as dotted lines for comparison.
A color version of this figure is available in the online journal.}
\label{fig:fNaD_line_strengths_ETG}
\end{center}
\end{figure*}

\subsection{Absorption Line Strengths}
\label{sec:line_ETG}

Spectral line strengths have been used to understand the physical processes governing 
the evolution of galaxies because they reflect the average surface gravities, effective 
temperatures and metal abundances of galaxies.

We first check whether the observed spectra match the models in specific index regions 
for \mgb, \fe\ and \hb\ by comparing the observed and expected model absorption strengths. 
The bottom three panels of Figure~\ref{fig:fNaD_ETG} show the \mgb\ (left, \fmgb), \fe\ 
(middle, \ffe) and \hb\ (right, \fhb) excess indices, which are calculated using the same 
procedure as in formula (1), against the \nad\ excess index \fnad. The contours of the 
early-type control sample (grey shaded) and all sample galaxies (dotted) are shown for 
comparison. The top panels of Figure~\ref{fig:fNaD_ETG} show the distributions of \fmgb, 
\ffe\ and \fhb, respectively. 

The early-type control sample and NEOs have very similar distributions of \fe\ and \hb\ 
excess indices, and these distributions  are Gaussian, unlike \fnad\ 
(see Figure~\ref{fig:fNaD}). However, there is mismatch in the \mgb\ excess index between 
the observed and expected model strengths, especially for ordinary early-type NEOs (red 
hashed histogram and red filled circles). This \mgb\ excess in ordinary early-type NEOs 
indicates that for these galaxies,  a mechanism to explain both the \nad\ and \mgb\ 
excesses is required, while \fmgb\ is much smaller than \fnad. For reference, the median 
values of \fnad, \fmgb, \ffe\ and \fhb\ for our early-type control sample are 0.06\,$\pm$\,0.03, 
0.07\,$\pm$\,0.08, $-$0.05\,$\pm$\,0.10 and 0.0\,$\pm$\,0.10, whereas those for ordinary 
early-type NEOs are 0.58\,$\pm$\,0.13, 0.26\,$\pm$\,0.11, $-$0.06\,$\pm$\,0.08 and 
$-$0.06\,$\pm$\,0.09, respectively.

\begin{figure}[t]
\begin{center}
\includegraphics[width=0.41\textwidth]{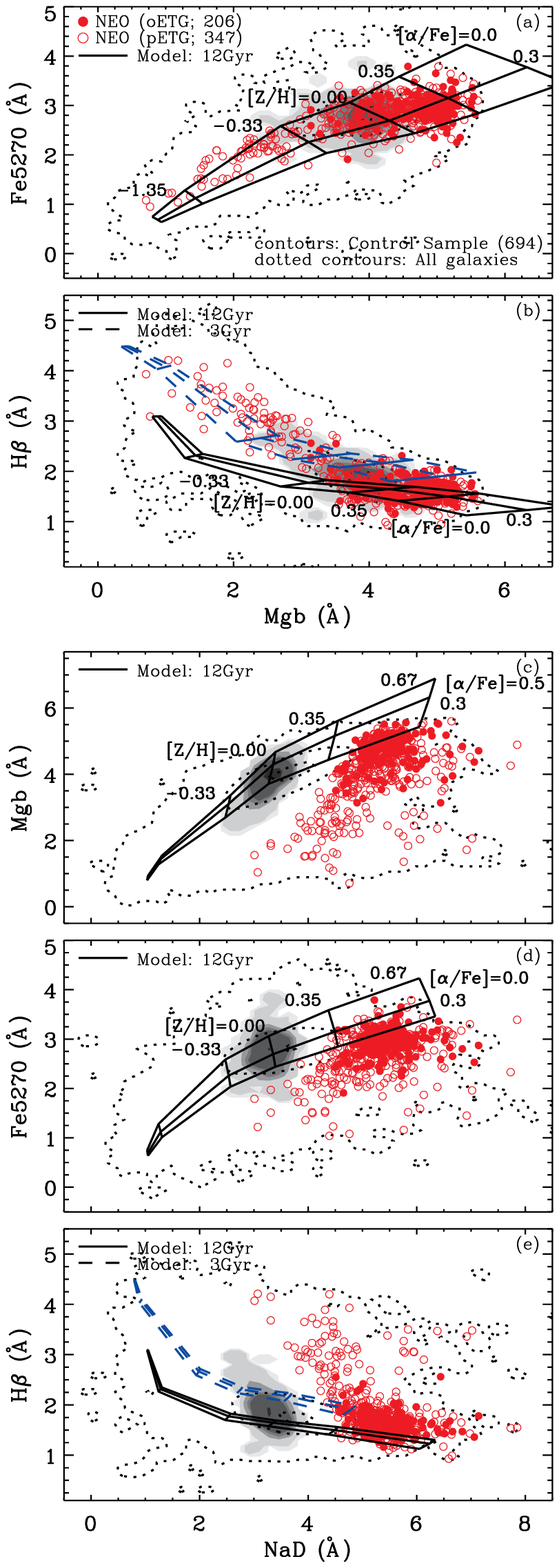}
\caption{Index measurements of our early-type control sample (grey shaded contours), 
ordinary (red filled circles) and peculiar (red open circles) early-type NEOs compared 
to the stellar population models of \citet{tmb03}. For comparison, the contours of all 
sample galaxies are shown as dotted lines. \mgb$-$\fe\ (a), \mgb$-$\hb\ 
(b), \nad$-$\mgb\ (c), \nad$-$\fe\ (d) and \nad$-$\hb\ (e) plots are shown.
A color version of this figure is available in the online journal.}
\label{fig:line_strengths_ETG}
\end{center}
\end{figure}

We now turn to the line strength itself. Strong Balmer absorption lines (e.g. \hb) betray the 
presence of young stars, while a combination of \mgb\ and \fe\ yields information about 
the mean metallicity of the population. In Figure~\ref{fig:fNaD_line_strengths_ETG}, we 
show a comparison of \fnad\ and the line strength indices of \nad\ (left), \mgb\ (middle) and 
\fe\ (right). The \nad\ excess index \fnad\ correlates strongly with the \nad\ line strength. 
This implies that NEOs are simply \nad\ strong galaxies. This may occur because the 
best-fit models for early-type NEOs generally expect \nad\ line strengths of around 3.0 to 
3.5\,\AA. Therefore, these galaxies have stronger \nad\ line strengths and higher \nad\ 
excesses. In contrast, the \mgb\ and \fe\ line strengths do not show significant 
correlations with \fnad\ due to contamination by peculiar early-type NEOs. Peculiar 
early-type NEOs show a wide range of \mgb\ and \fe\ values, often much lower values than 
those of the early-type control sample. We note that  some peculiar early-type NEOs with 
\fnad \,$\geqslant$\,1 have significantly weaker \mgb\ and \fe\ line strengths than the 
early-type control sample. These features are similar to late-type NEOs, as discussed in 
Section~\ref{sec:results_LTG} below.

In the top panels of Figure~\ref{fig:fNaD_line_strengths_ETG}, histograms of the \nad, 
\mgb\ and \fe\ line strengths are shown. Peculiar early-type NEOs had skewed distributions 
with a tail toward lower values for \mgb\ and \fe. Their \mgb\ and \fe\ strengths are similar to 
those of late-type galaxies. If some peculiar early-type galaxies have recently formed stars 
that are centrally concentrated, they would show similar line strength distributions to 
late-type galaxies because SDSS fibers encompass only central regions.

Line strength index diagrams comparing \nad, \mgb, \fe\ and \hb\ indices for the early-type 
control sample (grey shaded contours), ordinary (red filled circles) and peculiar (red open 
circles) early-type NEOs based on the stellar population models of Thomas, Maraston \& 
Bender (2003) are shown in Figure~\ref{fig:line_strengths_ETG}. Contours of all sample 
galaxies are also shown as dotted lines for comparison. For the \mgb$-$\fe\ and \mgb$-$\hb\  
plots ((a) and (b), Figure~\ref{fig:line_strengths_ETG}), our sample galaxies follow the 
models reasonably well. The majority of (ordinary) early-type NEOs are old, consistent with 
an isochrone of 12\,Gyr, metal-rich above the solar value, and an over abundance in 
$\alpha$-elements, which are the general characteristics of massive early-type galaxies 
\citep{r88,gea90,wfg92,tetal00,tetal05}. However, some peculiar early-type galaxies have 
stronger \hb\ line strengths than the early-type control sample, which is evidence of recent 
star formation \citep[see e.g.][]{yetal05,ketal07,jetal07,jetal09,setal10,cetal11}.

In the \nad$-$index plots ((c)$-$(e), Figure~\ref{fig:line_strengths_ETG}), the sample 
galaxies deviate significantly from the model grids. They show enhancement of \nad\ with 
respect to \mgb,  \fe\ and \hb. This behavior indicates that the \nad\ line strengths of some 
galaxies, especially at higher \nad\ strength regions (\nad\,$\geqslant$\,4.0\,\AA), are 
difficult to replicate using stellar population models. This brings stellar population models 
in question, consistent with previous suggestions (e.g. IMF variation and/or overabundance 
of [Na/Fe]), or indicates that the \nad\ excess may be caused by non-stellar components 
(e.g. ISM and/or dust).

\begin{figure*}[t]
\begin{center}
\includegraphics[width=0.85\textwidth]{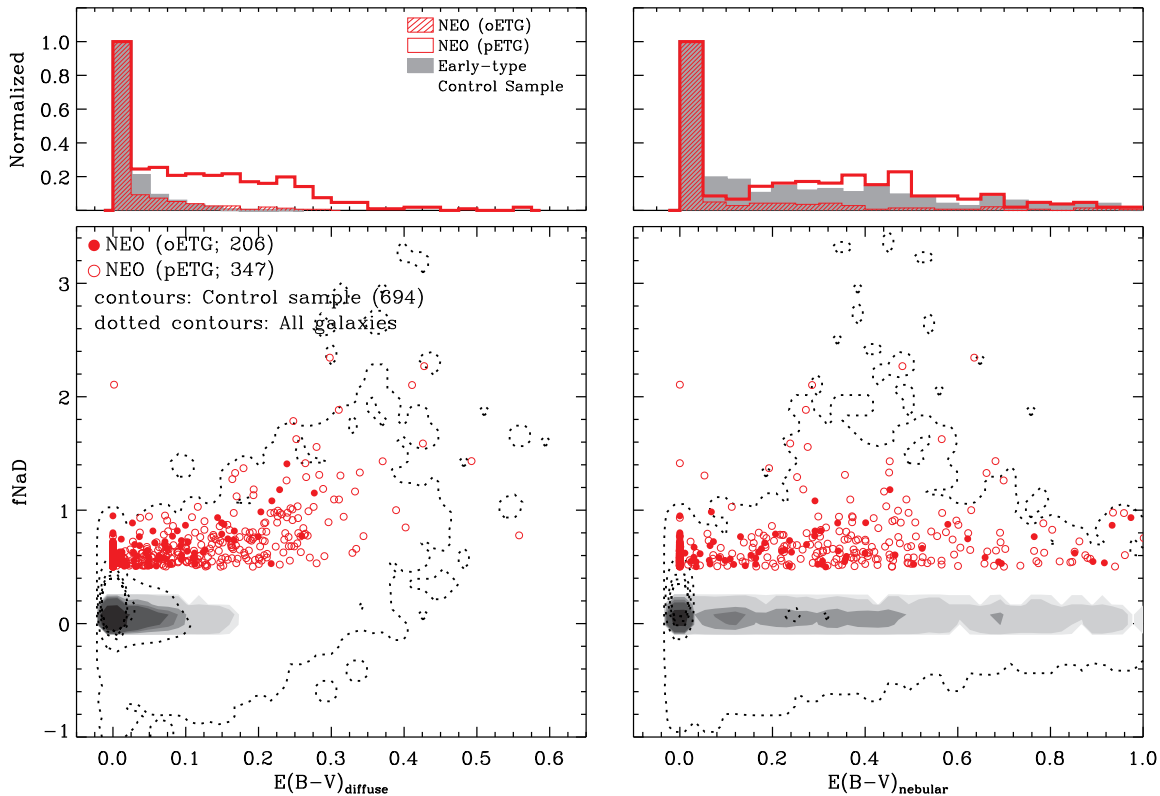}
\caption{\textit{Top}: Distribution of E(B-V) values for the early-type control sample 
(grey filled), ordinary (red hashed) and peculiar (red solid) early-type \nad\ excess 
objects. 
\textit{Bottom}: \fnad$-$E(B-V) plot for the early-type control sample (grey shaded 
contours), ordinary (red filled circles) and peculiar (red open circles) early-type 
NEOs. The contours of the sample galaxies are shown as dotted lines for comparison.
A color version of this figure is available in the online journal.}
\label{fig:ebv_ETG}
\end{center}
\end{figure*}

\subsection{Impact of Interstellar Extinction}
\label{sec:ebv_ETG}

It is well known that the \nad\ index is influenced by the ISM, and dust lanes provide 
additional absorption in this line. Some studies have reported that dust extinction correlates 
with \nad\ line strength \citep[see e.g.][]{ba86, cetal10, ppb12}. We therefore consider the 
impact of interstellar extinction on the \nad\ line index to check whether the observed \nad\ 
line strength of early-type NEOs can be explained by the ISM and/or dust. 

The OSSY catalogue provides two distinct \ebv\ reddening measurements: (1) \ebvs ,
corresponding to a diffuse ``screen'' dust component that affects the entire spectrum and  
(2) \ebvg , associated with a ``nebular'' component that impacts only the ionized-gas 
emission, which is known to be generally related to dust features \citep{setal06}. \ebvs\ 
values can be reasonably estimated if the (good-quality) continuum spectra match well 
with those generated by stellar population models, as is the case for our sample galaxies, 
whereas \ebvg\ values can be constrained only in the presence of nebular emissions. 
Moreover, \ebvg\ values depend on our assumption, of an intrinsic Balmer decrement 
(for more details see OSSY).

In the left panel of Figure~\ref{fig:ebv_ETG}, we show a comparison of \fnad\ and the dust 
extinction values of \ebvs\ for our sample galaxies using histograms. Early-type control 
sample (grey filled) and our ordinary early-type NEOs (red hashed) have nearly identical 
distributions with very little dust extinction from a diffuse screen dust component. In fact, 
the uncertainties of our \ebvs\ measurements are generally so small (less than 0.01 for 
spectra with an S/N\,$>$\,20, our minimum data quality threshold; see 
Section~\ref{sec:sample} and Table~\ref{tab:sample_crit}) that our \ebvs\ measurements 
are consistent with little or no diffuse dust in our ordinary early-type NEOs, which suggests 
that their \nad\ excess may be related to specific aspects of their stellar populations (e.g. 
different abundance patterns). In contrast, roughly 50\,\% of peculiar early-type NEOs have 
significant \ebvs\ values greater than 0.1. This implies that our peculiar early-type NEOs are 
generally dustier systems than ordinary early-type NEOs and the early-type control sample. 
This finding is expected because we classified dust lane early-type NEOs as peculiar 
galaxies.

A comparison of \fnad\ and \ebvg\ values for our early-type NEOs and early-type control 
sample galaxies is provided in the right panel of Figure~\ref{fig:ebv_ETG}. The \ebvg\ 
values, especially those of the early-type control sample and peculiar early-type NEOs, 
have skewed distributions with tails toward higher values. The extent of such tails most 
likely reflects the different incidence of ionized-gas emissions in our early-type 
subsamples (see Table~\ref{tab:bpt_ETG}), as expected given the connection between 
dust features and emission-line regions \citep{setal06}, whereas the overall similarity of 
the distributions indicates that nebular dust is not related to \nad\ excess.


\begin{figure}[t]
\begin{center}
\includegraphics[width=0.48\textwidth]{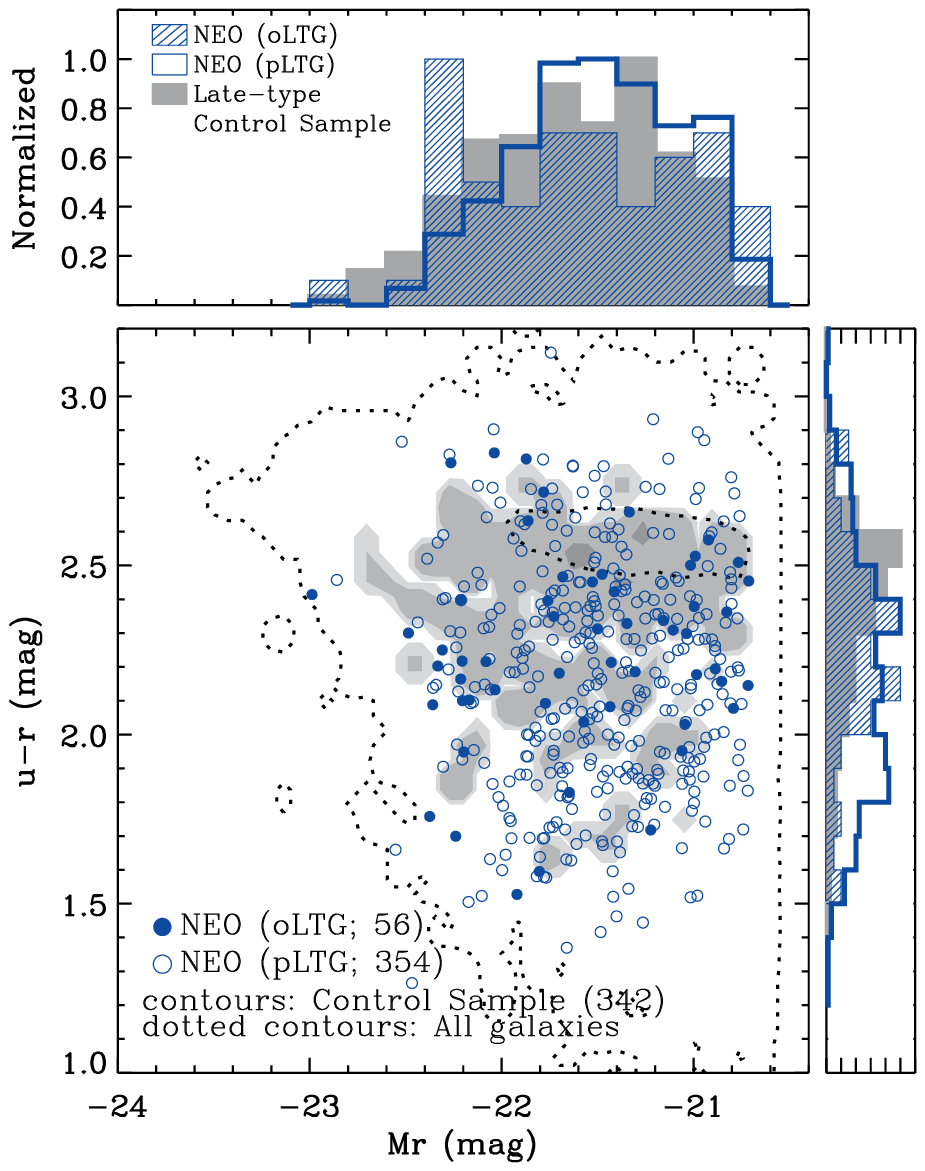}
\caption{Same as Figure~\ref{fig:cmr_ETG} but for the late-type control sample 
(grey filled histogram and shaded contours), ordinary (blue hashed histogram 
and filled circles, oLTGs) and peculiar (blue solid histogram and open circles, 
pLTGs) late-type \nad\ excess objects.
A color version of this figure is available in the online journal.}
\label{fig:cmr_LTG}
\end{center}
\end{figure}

\begin{figure}[t]
\begin{center}
\includegraphics[width=0.49\textwidth]{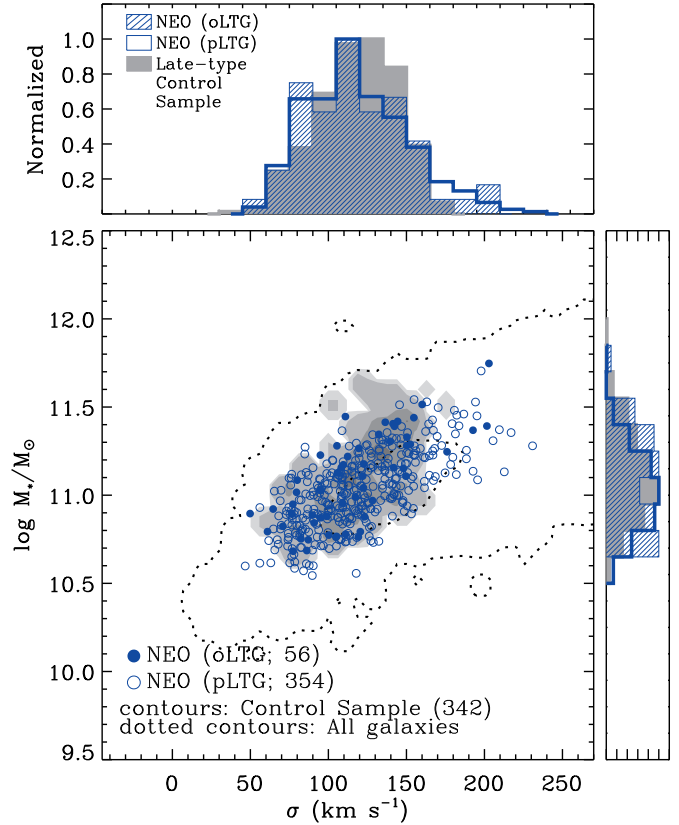}
\caption{Same as Figure~\ref{fig:vd_ETG} but for the late-type control sample 
(grey filled histogram and shaded contours), ordinary (blue based histogram 
and filled circles, oLTGs) and peculiar (blue solid histogram and open circles, 
pLTGs) late-type \nad\ excess objects.
A color version of this figure is available in the online journal.}
\label{fig:vd_LTG}
\end{center}
\end{figure}

\section[]{Properties of late-type\\ \nad\ excess objects}
\label{sec:results_LTG}

\subsection{Color-Magnitude Relation}
\label{sec:cmr_LTG}

It is now common practice to divide galaxy populations on a color-magnitude diagram. 
Late-type galaxies typically have bluer colors than early-type galaxies and seem to reside 
in the blue cloud or green valley.

In Figure~\ref{fig:cmr_LTG}, we present u$-$r CMR (bottom-left) with M$_{r}$ (top) and 
u$-$r (bottom-right) histograms of our late-type control sample (grey shaded contours and 
grey filled histogram), ordinary late-type NEOs (blue filled circles and hashed histogram, 
oLTGs) and peculiar late-type NOEs (blue open circles and solid histogram, pLTGs). A 
cursory inspection of the plot shows that our late-type NEOs have similar overall ranges of 
absolute $r$-band magnitudes and slightly bluer u$-$r optical color distributions than the 
late-type control sample, in contrast to early-type NEOs, which are more luminous and 
optically redder than the early-type control sample. Our late-type NEOs also have a 
narrower range of M$_r$, but a wider range of colors, than our early-type NEOs.  We note 
that the difference in peak position of ordinary late-type NEOs may be due to limited number 
statistics. It is also worth noting that our late-type sample contains a number of red spiral 
galaxies in the green valley and even in the red sequence.

\subsection{Velocity Dispersion and Stellar Mass}
\label{sec:vd_LTG}

The effective velocity dispersion and stellar mass of late-type galaxies are also derived 
using formulas (1) and (2), respectively. It is important, however, to recognize that these 
formulas we derived mainly for early-type galaxies, which may lead to  measurement 
errors when applying them to late-type galaxies.

Figure~\ref{fig:vd_LTG} shows the relationship between the velocity dispersion and 
stellar mass. The control sample (grey shaded contours) and late-type NEOs (blue filled 
and open circles) have very similar velocity dispersions and stellar mass distributions (or 
slightly lower velocity dispersions and smaller stellar masses), unlike the early-type case 
for which there was a correlation between \nad\ line strengths and velocity dispersions.
We also plot the distributions of velocity dispersion and stellar mass of the late-type control 
sample (grey filled), ordinary late-type NEOs (blue hashed) and peculiar late-type NEOs 
(blue solid). We find again that the velocity dispersion and stellar mass distributions of our 
late-type NEOs  are virtually indistinguishable from those of their control counterparts.

\begin{figure*}[t]
\begin{center}
\includegraphics[width=0.95\textwidth]{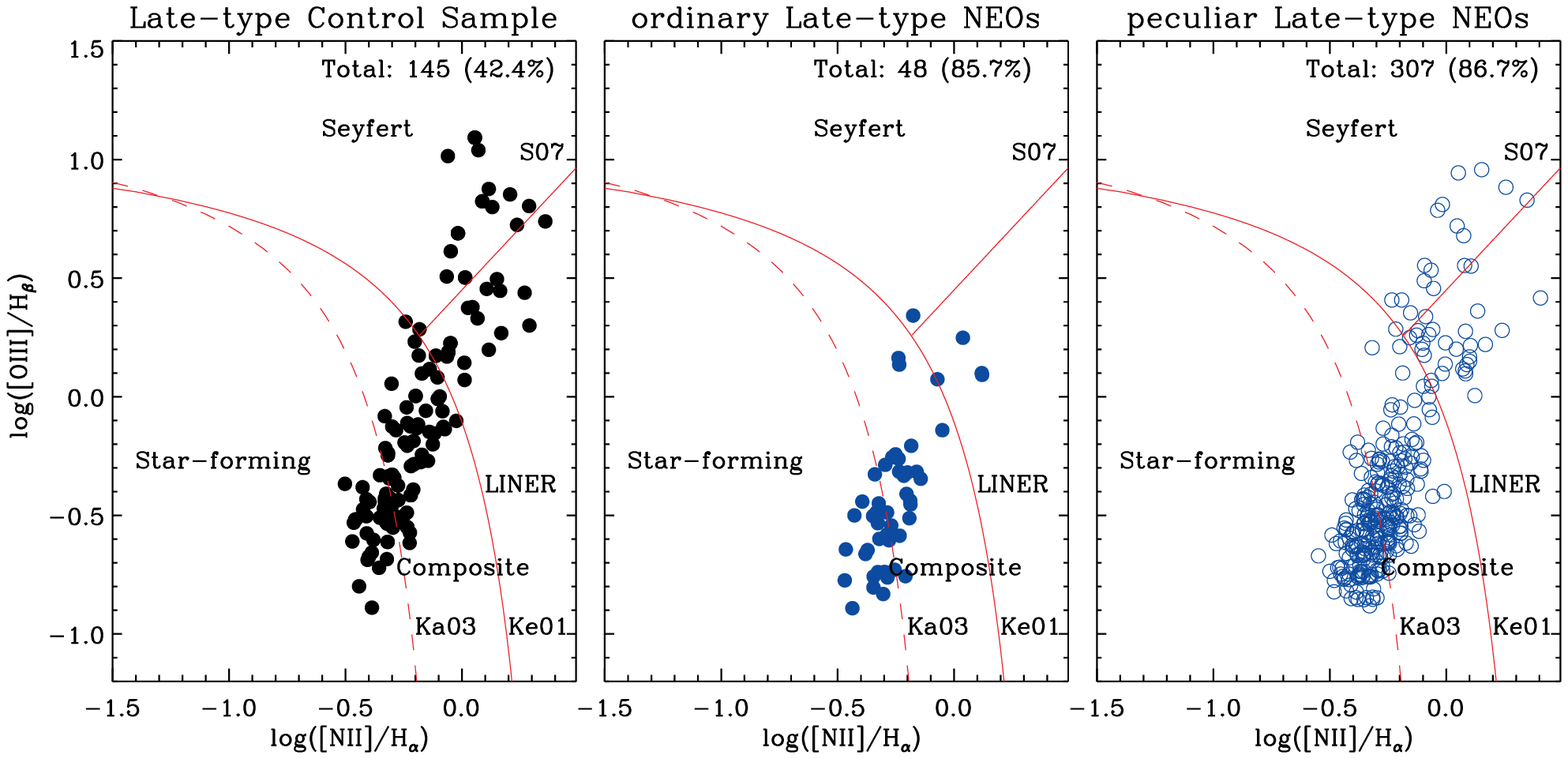}
\caption{ Same as Figure~\ref{fig:bpt_ETG} but for the late-type control sample 
(black filled circles, left), ordinary (blue filled circles, middle) and peculiar (blue 
open circles, right) late-type \nad\ excess objects.
A color version of this figure is available in the online journal.}
\label{fig:bpt_LTG}
\end{center}
\end{figure*}

\subsection{Emission Line Diagnostics: \\ Star Formation and AGN Activity}
\label{sec:bpt_LTG}

We begin our spectroscopic analysis by exploring the emission line diagnostic of our 
late-type galaxies using BPT diagrams (see section~\ref{sec:bpt_ETG} above). 

BPT diagrams for our late-type galaxies are shown in Figure~\ref{fig:bpt_LTG}. Note again 
that we only plot galaxies where [N\,{\small II}], H$\alpha$, [O\,{\small III}] and H$\beta$ 
lines are detected with A/N\,$\geqslant$\,3. Black filled circles (left) represent the late-type 
control sample, while ordinary and peculiar late-type NEOs are indicated by blue filled 
circles (middle) and blue open circles (right), respectively. To separate galaxies, we use 
the same demarcation lines as we used for early-type galaxies (see 
Section~\ref{sec:bpt_ETG} and Figure~\ref{fig:bpt_ETG}). Table~\ref{tab:bpt_LTG} 
summarizes the star-forming, composite and AGN properties of our late-type galaxies.

Interestingly, almost 90\,\% of late-type NEOs show significant emission lines in contrast to 
early-type NEOs, which are virtually free of emission lines. Star forming galaxies, in 
particular, are more common in both ordinary and peculiar late-type NEOs. The fractions 
of late-type NEOs classified as star-forming and composite are larger by factors of 4 and 2, 
respectively, than those of the late-type control sample, whereas the fractions classified as 
Seyfert or LINER are slightly lower. Therefore, it appears that the \nad\ excess in late-type 
galaxies may be related to phenomena associated with star formation.

\begin{table*}[t]
 \begin{center}
  \caption{Results of Spectral Line Classification for Late-type Galaxies}
  \begin{tabular}{llrrrr}
  \hline \hline
 \multicolumn{1}{c}{Classification} & & late-type control sample & late-type NEOs & oLTGs$^a$ & pLTGs$^b$ \\
 \hline
 Quiescent            &           &  57.6\% (197)  &  13.4\% (55)    & 14.3\% (8)    & 13.3\% (47)\\
 Strong emission &           &  42.4\% (145)  &  86.6\% (355)  & 85.7\% (48)  & 86.7\% (307)\\
       & Star-forming            &  12.0\% (41)    &  41.0\% (168)  & 39.3\% (22)  & 41.2\% (146)\\
       & Composite               &  18.7\% (64)    &  33.4\% (137)  & 37.5\% (21)  & 32.8\% (116)\\
       & Seyfert                      &    5.6\% (19)     &    4.2\% (17)     &   1.8\% (1)    &   4.5\% (16)\\
       & LINER                       &    6.1\% (21)     &    8.0\% (33)     &   7.1\% (4)    &   8.2\% (29)\\ 
  \hline
 \end{tabular}
 \label{tab:bpt_LTG}
 \end{center}
 \bf{~~Notes.}
 \tablenotetext{a}{ordinary late-type \nad\ excess objects}
 \tablenotetext{b}{peculiar late-type  \nad\ excess objects}
\end{table*}

\begin{figure*}[t]
\begin{center}
\includegraphics[width=0.85\textwidth]{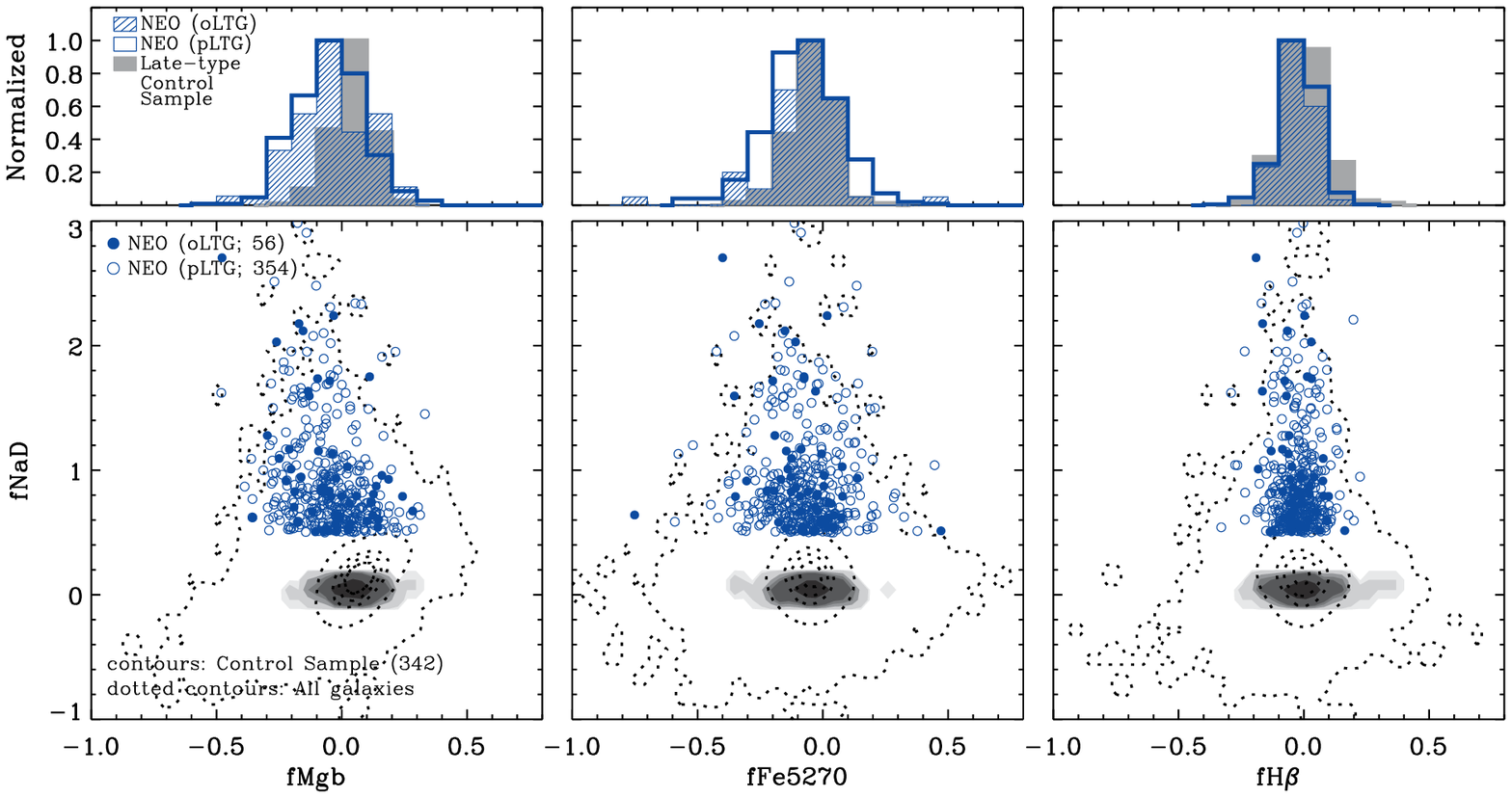}
\caption{Same as Figure~\ref{fig:fNaD_ETG} but for the late-type control sample 
(grey filled histogram and shaded contours), ordinary (blue hashed histogram and 
filled circles, oLTGs) and peculiar (blue solid histogram and open circles, pLTGs) 
late-type \nad\ excess objects.
A color version of this figure is available in the online journal.} 
\label{fig:fNaD_LTG}
\end{center}
\end{figure*}

\begin{figure*}[t]
\begin{center}
\includegraphics[width=0.85\textwidth]{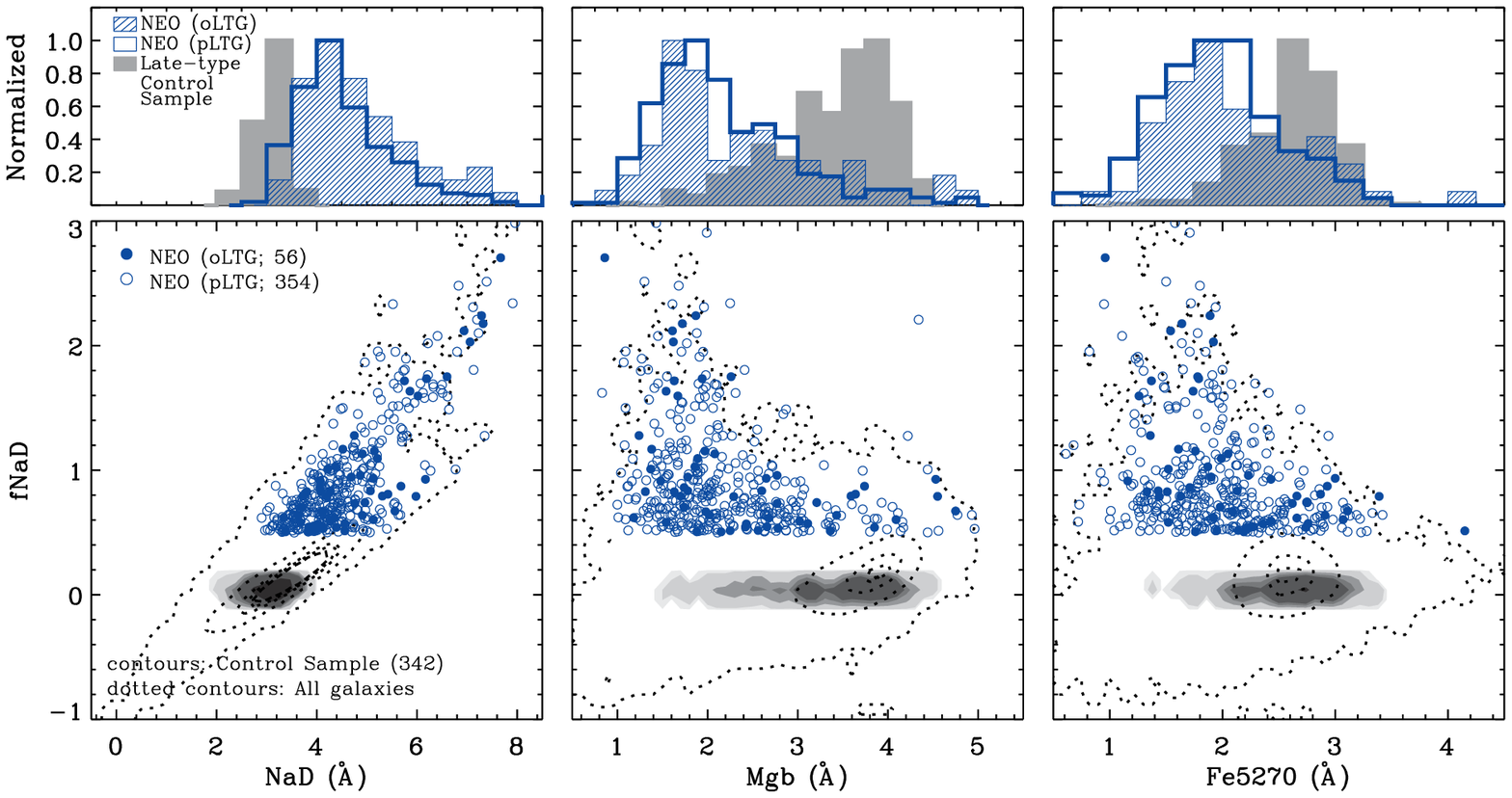}
\caption{Same as Figure~\ref{fig:fNaD_line_strengths_ETG} but for the late-type 
control sample (grey filled histogram and shaded contours), ordinary (blue hashed 
histogram and filled circles, oLTGs) and peculiar (blue solid histogram and open 
circles, pLTGs) late-type \nad\ excess objects.
A color version of this figure is available in the online journal.} 
\label{fig:fNaD_line_strengths_LTG}
\end{center}
\end{figure*}

\subsection{Absorption Line Strengths}
\label{sec:line_LTG}

We check again the excesses of \mgb, \fe\ and \hb\ in late-type galaxies by comparing the 
line strengths between the observations and best-fit model. The bottom panels of 
Figure~\ref{fig:fNaD_LTG} show the \mgb\ (left, \fmgb), \fe\ (middle, \ffe) and \hb\ (right, \fhb) 
excess indices, which are calculated using the same procedure as in formula (1), as a 
function of \fnad. Blue filled and open circles represent  ordinary and peculiar late-type 
NEOs, respectively. The dotted contours indicate the distribution of all sample galaxies, 
while the grey shaded contours indicate the distribution of late-type control sample galaxies. 
Note that most late-type NEOs have negative \fmgb\ values, while early-type NEOs tend to 
have positive \fmgb\ values (see Figure~\ref{fig:fNaD_ETG}). This is likely related to 
enhanced star formation in late-type NEOs, as discussed in Section~\ref{sec:bpt_LTG} 
above. Late-type NEOs have \ffe\ and \fhb\ distributions very similar with those of the 
late-type control sample.

\begin{figure}[t]
\begin{center}
\includegraphics[width=0.425\textwidth]{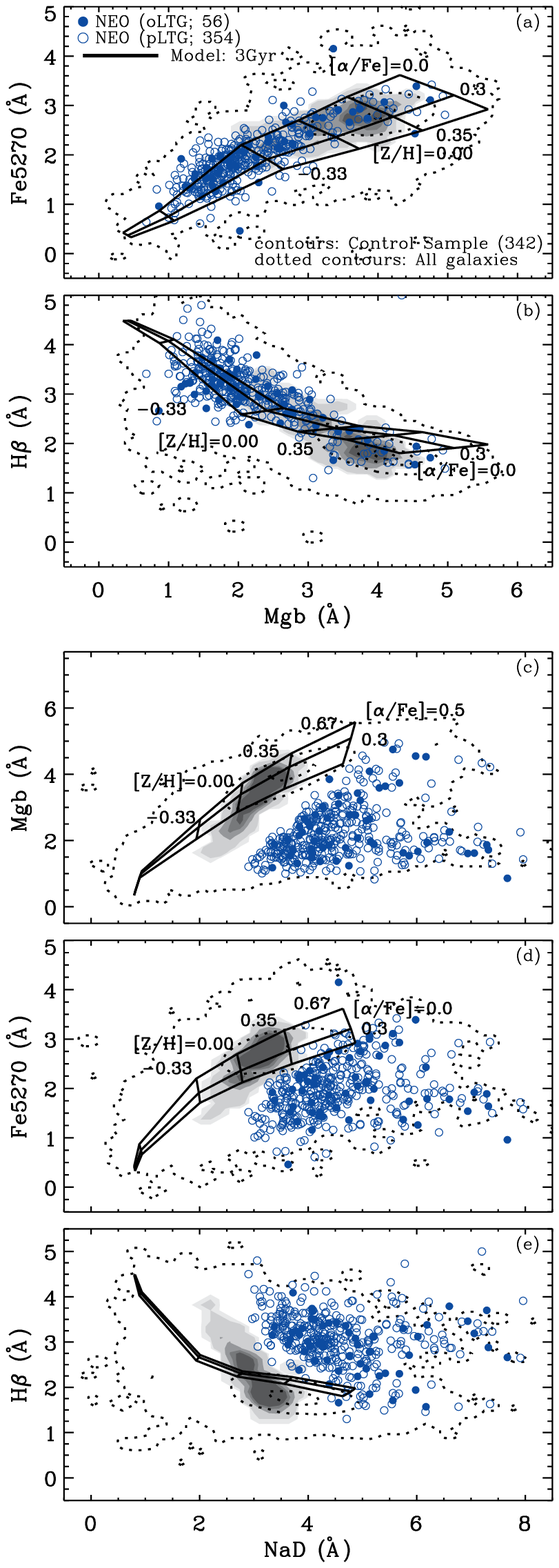}
\caption{Same as Figure~\ref{fig:line_strengths_ETG} but for the late-type control 
sample (grey filled histogram and shaded contours), ordinary (blue hashed 
histogram and filled circles, oLTG) and peculiar (blue solid histogram and open 
circles, pLTG) late-type \nad\ excess objects.
A color version of this figure is available in the online journal.} 
\label{fig:line_strengths_LTG}
\end{center}
\end{figure}

In the top panels of Figure~\ref{fig:fNaD_LTG}, we also plot the distributions of \fmgb, \ffe\ 
and \fhb. We confirm that the peak positions of the \fmgb\ histograms for the late-type 
control sample (0.05) and late-type NEOs ($-$0.05) are slightly different. In the cases of 
\ffe\ and \fhb, the distributions of late-type NEOs are virtually indistinguishable from those 
of their control counterparts (less than 0.03 difference on average), but late-type NEOs 
show wider deviations in \ffe\ than the late-type control sample. For reference, the median 
values of \fnad, \fmgb, \ffe\ and \fhb\ for our late-type control sample are 0.04\,$\pm$\,0.03, 
0.05\,$\pm$\,0.08, $-$0.05\,$\pm$\,0.09 and 0.00\,$\pm$\,0.09, whereas those for ordinary 
late-type NEOs are 0.83\,$\pm$\,0.52, $-$0.04\,$\pm$\,0.15, $-$0.08\,$\pm$\,0.16 and 
$-$0.03\,$\pm$\,0.07, respectively.

Figure~\ref{fig:fNaD_line_strengths_LTG} shows the line strengths of  \nad\ (left), \mgb\ 
(middle) and \fe\ (right) versus \fnad.  We again see a correlation between \fnad\ and \nad\ 
line strengths, as discussed in Section~\ref{sec:line_ETG}. In contrast, the \mgb\ and \fe\ 
line strengths do not show any significant correlations with \fnad, but late-type NEOs (blue 
filled and open circles) tend to have weaker \mgb\ and \fe\ line strengths than the late-type 
control sample (grey shaded contours) in contrast to the majority of early-type NEOs, which 
have stronger \mgb\ and \fe\ line strengths. In the top panels of 
Figure~\ref{fig:fNaD_line_strengths_LTG}, the \nad, \mgb\ and \fe\ line strength distributions 
of late-type NEOs and those of their control-sample counterparts are shown; late-type 
NEOs have stronger \nad\ line strengths but weaker \mgb\ and \fe\ line strengths than their 
control-sample counterparts.

In Figure~\ref{fig:line_strengths_LTG}, we show the absorption line measurements of our 
late-type galaxies compared to the stellar population models of \citet{tmb03}. The upper 
two panels of Figure~\ref{fig:line_strengths_LTG} indicate that our late-type galaxies are 
very well represented by a coeval (3\,Gyr old) sequence of models with various metallicities, 
while the lower three panels associated with the \nad\ line strength show that our data 
deviate from the model grids as in the case of early-type NEOs (see 
Figure~\ref{fig:line_strengths_ETG}). One notable difference is that most late-type NEOs 
tend to have weaker \mgb\ and \fe\ line strengths with enhanced \hb\ strengths, which is 
further evidence of recent star formation, even though this may not be directly connected to 
\nad\ excess. 

\begin{figure*}[t]
\begin{center}
\includegraphics[width=0.85\textwidth]{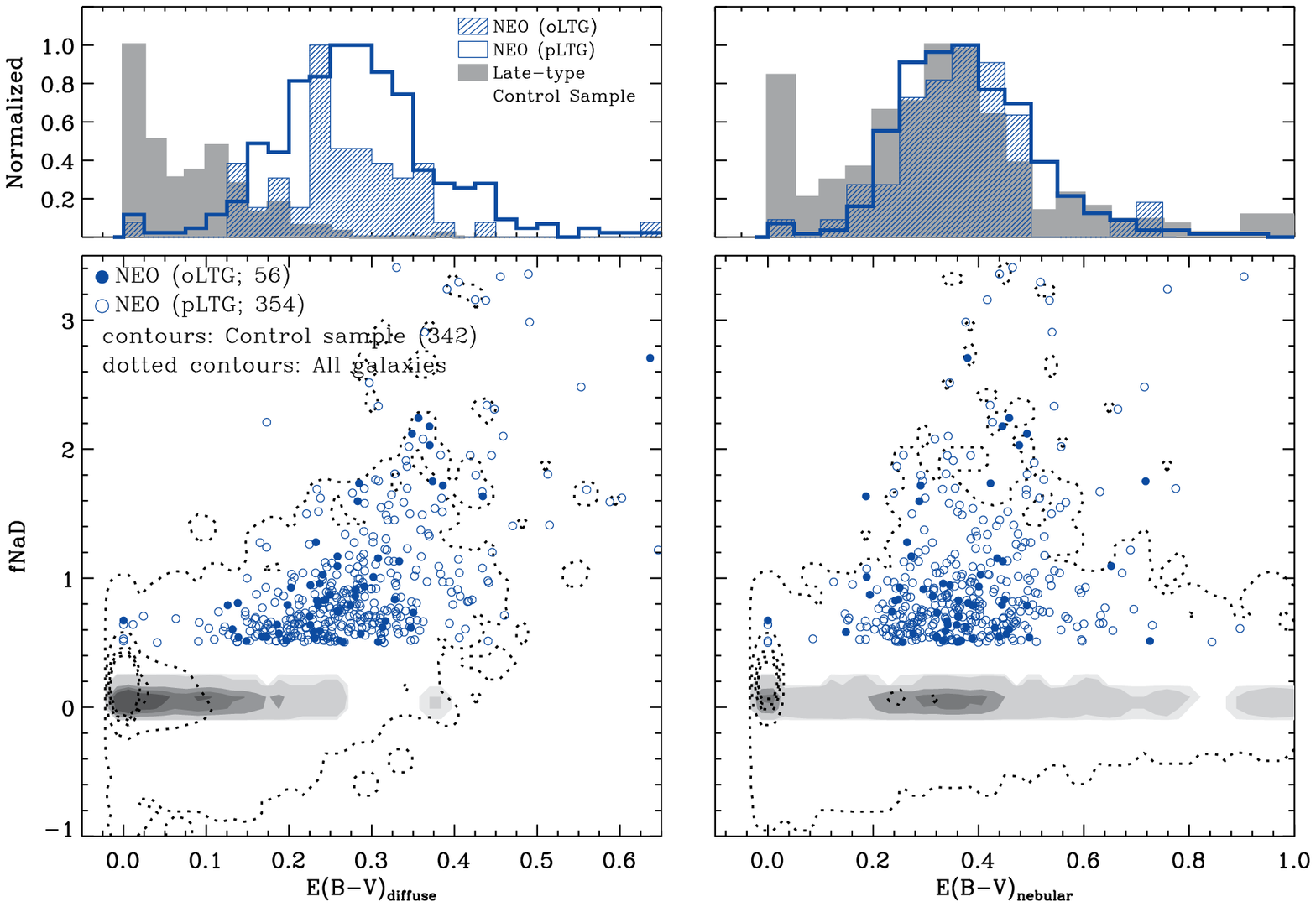}
\caption{Same as Figure~\ref{fig:ebv_ETG} but for the late-type control 
sample (grey filled histogram and shaded contours), ordinary (blue hashed 
histogram and filled circles, oLTGs) and peculiar (blue solid histogram and open 
circles, pLTGs) late-type \nad\ excess objects. 
A color version of this figure is available in the online journal.}
\label{fig:ebv_LTG}
\end{center}
\end{figure*}

\subsection{Impact of Interstellar Extinction}
\label{sec:ebv_LTG}

We consider again the impact of interstellar extinction by comparing the \nad\ excess 
index \fnad\  and the dust extinction values of \ebvs\ and \ebvg. We note that both ordinary 
and peculiar late-type NEOs stand out from the late-type control sample galaxies due to 
their much larger diffuse screen dust component values (left panel of 
Figure~\ref{fig:ebv_LTG}), whereas late-type control sample and late-type NEOs share 
very similar distributions of \ebvg\ relating to the nebular component (right panel of 
Figure~\ref{fig:ebv_LTG}). This suggests not only that the ISM plays an important role in 
explaining the \nad\ excess found in late-type NEOs, but also that the ISM contribution to 
the \nad\ lines relates only to the diffuse dust component, which also reinforces our 
previous interpretation of \nad\ excess in ordinary (dust-free) early-type NEOs. We also 
further note that the lack of a connection between the nebular dust component and \nad\ 
excess is not completely unexpected if we consider that emission-line regions can be 
quite clumpy, so that only a small fraction of stellar light encompassed by the SDSS spectra 
is covered.
 
According to some previous studies \citep[see e.g.][]{ba86, cetal10, ppb12}, the connection 
between diffuse screen dust extinction and \nad\ line strength can be valuated quantitatively. 
Adopting the recent calibration of \citet{ppb12} for the impact of interstellar absorption on 
\nad\ line strength (see their equation 9), the average \ebvs\ value of $\sim0.3$ magnitudes 
(see Figure~\ref{fig:ebv_LTG}) observed in our late-type NEOs would be sufficient to explain 
the main peak shift of $\sim1$\,\AA\ between the distributions of \nad\ line strengths for 
late-type NEOs and that corresponding to the late-type control sample 
(see Figure~\ref{fig:fNaD_line_strengths_LTG}). This calibration also allows us to confirm 
that the impact of the ISM on \nad\ line strengths in our ordinary early-type NEOs is negligible, 
even when using a value of 0.01\,\AA\  as an upper limit for \ebvs , corresponding to typical 
uncertainties associated with this quantity (see Section~\ref{sec:ebv_ETG}).  

\section{Discussion}
\label{sec:discussion}

A number of recent papers exploring Na features that are known to depend on surface 
gravity have indicated that there is IMF variation among galaxies. To investigate the nature 
of NEOs, we selected \nad\ excess candidates in the redshift range 
0.00\,$\leqslant$\,$z$\,$\leqslant$\,0.08 from SDSS DR7 by adopting a new index, \fnad. 
This index quantifies \nad\ excess by comparing the observed spectrum with the best-fit 
model spectrum. Roughly 8\,\% (1\,603/20\,571, \fnad\,$\geqslant$\,0.5) of galaxies in the 
sample were classified as NEOs. These galaxies where then identified through direct 
visual inspection of SDSS images, resulting in a sample of 553 early-type and 410 
late-type NEOs. An important goal of this paper was a systematic comparison of the 
properties of NEOs and those of a control sample based on homogeneous data sets. Note 
that only 0.5\,\% of galaxies (17/3\,461) in the \nad\ excess range 
0.0\,$\leqslant$\,\fnad\,$\leqslant$\,0.1 had high velocity dispersions greater than 
250 km\,s$^{-1}$. This implies that most high velocity dispersion galaxies are classified as 
NEOs. The results presented in Sections~\ref{sec:results_ETG} and \ref{sec:results_LTG} 
indicate that the origin of \nad\ excess may differ considerably depending on galaxy 
morphology.

The majority of early-type NEOs are optically redder, more luminous,  more massive and 
more likely to be high velocity dispersion systems than early-type control sample 
counterparts. Furthermore, they are stronger in \mgb\ and \fe\  than their control-sample 
counterparts, which have roughly solar abundance values (e.g. [Z/H]\,$\sim$\,0 and 
[$\alpha$/Fe]\,$\sim$\,0). 

\begin{figure}[t]
\begin{center}
\includegraphics[width=0.48\textwidth]{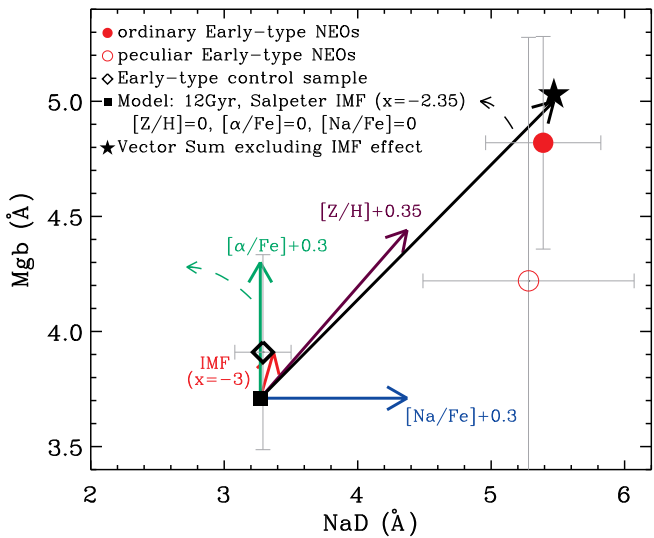}
\caption{Response of \nad\ and \mgb\ line indices to changes in the abundance patterns 
([$\alpha$/Fe], [Z/H] and [Na/Fe]) and the IMF. The mean observed \nad\ and \mgb\ line 
strengths of the early-type control sample, ordinary and peculiar early-type NEOs are shown 
by the open diamond, red filled circle and red open circle, respectively. Black filled star 
represents the vector sum excluding the IMF effect.
A color version of this figure is available in the online journal.}
\label{fig:model}
\end{center}
\end{figure}

Mean observed \mgb\ and \nad\ line strengths of the early-type control sample (black open 
diamond), ordinary early-type NEOs (red filled circle) and peculiar early-type NEOs (red 
open circle) with a specific stellar population model (black filled square) are shown in 
Figure~\ref{fig:model}. The errors correspond to standard deviations. A galaxy with an age 
of 12\,Gyr and solar metallicity (e.g. [Z/H]\,=\,0, [$\alpha$/Fe]\,=\,0 and [Na/Fe]\,=\,0) with a 
Salpeter IMF \citep{s55} based on the models of \citet[][see also Lee, Worthey \& Dotter 2009; 
\citealt{vc12}]{tmb03} will have \nad\ and \mgb\ line strengths of 3.27 and 3.71\,\AA\ (black 
filled square), respectively. These model values are consistent with the observed values of 
early-type control samples (open diamond).

In contrast, the observed line strengths of early-type NEOs were notably different from the 
model values (see Figure~\ref{fig:line_strengths_ETG}). Given that the \nad\ index responds 
more strongly to [Na/Fe] than the IMF and is even influenced by the ISM, one might be 
tempted to disregard this mismatch. However, it is important, because it provides information 
about sodium abundance [Na/Fe]. To find stellar population models that reproduce the 
observed line strengths, we explored various factors that may increase the \nad, \mgb\ and 
\fe\ line strengths based on the models of 
\citet[][hereafter TMB, 12\,Gyr models]{tmb03},  
\citet[][hereafter LWD, 12\,Gyr models]{lwd09} and 
\citet[][hereafter CV, 13.5\,Gyr models]{cv12}.

$\bullet$\,$\alpha$-enhancement effect (green solid and dashed arrows in 
Figure~\ref{fig:model}):  The \mgb\ index, which is a well-known tracer of $\alpha$-elements, 
changes by 0.59\,\AA\ when [$\alpha$/Fe] varies by +0.3 dex (TMB \& LWD, green solid 
arrow), while \fe\ decreases about 0.37\AA\ with an increasing [$\alpha$/Fe] (TMB \& LWD). 
The effect of the $\alpha$-enhancement on \nad, however, is not clear. According to the 
models of TMB, the \nad\ feature increases slightly with $\alpha$-enhancement (see 
Figures~\ref{fig:line_strengths_ETG}c or \ref{fig:line_strengths_LTG}c), whereas in the 
models of CV \& LWD, the \nad\ line index has weakened about 0.55\,\AA\ on the 
$\alpha$-enhancement side ([$\alpha$/Fe]\,=\,0.3). The green and black dashed curved 
arrows in Figure~\ref{fig:model} represent this tendency. This means that 
$\alpha$-enhancement is not conducive to increasing \nad\ line strength. It is crucial, 
however, to reproduce the observed \mgb\ line strength, and it is compatible with our general 
understanding of massive early-type galaxies 
\citep[see e.g.][]{r88,gea90,wfg92,tetal00,tetal05}.

$\bullet$\,Metal effect (violet solid arrow in Figure~\ref{fig:model}): The \mgb\ and \fe\ indices 
increase about 0.73 and 0.52\,\AA\ when [Z/H] varies from 0.0 to 0.35 dex. An increase in [Z/H] 
by 0.35 dex also causes an increase in the strength of \nad\ by about 1.10\,\AA\ (TMB).

$\bullet$\,IMF effect (red solid arrow in Figure~\ref{fig:model}): The differences in \mgb\ and 
\nad\ line strengths between the Salpeter and x\,=\,$-$3 bottom-heavy IMF are only 0.2 and 
0.1\,\AA, respectively. This implies that IMF variation has a limited impact on \mgb\ and \nad\ 
(CV). 

$\bullet$\,Na-enhancement effect  (blue solid arrow in Figure~\ref{fig:model}):  The \nad\ line 
index is known to be particularly sensitive to [Na/Fe]. At [Na/Fe]\,=\,0.3, \nad\ increases about 
1.1\,\AA\ (CV).

Given that massive early-type galaxies are known to have an overabundance of 
$\alpha$-elements and super-solar total metallicity, it is natural to assume that (ordinary) 
early-type NEOs have high [$\alpha$/Fe] and [Z/H] abundances. Furthermore, our early-type 
NEOs show strong \mgb\ and \fe\ line strengths, as described above and in 
Section~\ref{sec:results_ETG}. Furthermore, the observed \mgb\ (4.82\,\AA) and \fe\ (2.98\,\AA) 
line strengths and the peak value of u$-$r color (2.75 mag) of ordinary early-type NEOs are 
consistent with the models of TMB (4.29\,\AA \,$\leqslant$\,\mgb \,$\leqslant$\,5.16\,\AA\ and  
2.69\,\AA \,$\leqslant$\,\fe \,$\leqslant$\,3.15\,\AA\ when [$\alpha$/Fe]\,=\,0.3 and 
0.0\,$\leqslant$\,[Z/H]\,$\leqslant$\,0.35) and \citet{yi03} (u$-$r\,=\,2.74 when Z\,=\,0.04). 
However, the observed \nad\ line strength cannot be reproduced with these two assumptions, 
and the bottom-heavy IMF does not do contribute substantially to increase the \nad\ strength. 
We therefore evaluated the hypothesis of enhanced Na abundance.

\begin{table*}[t]
 \begin{center}
  \caption{A sample of the catalogue of \nad\ excess objects.}
  \begin{tabular}{ccccccccccc}
  \hline \hline
 \multicolumn{1}{c}{SDSS object id} & RA  & DEC & Redshift & M$_r$ & \fnad\ & \nad$^a$ & \fmgb\ & \mgb$^a$  & Morphology & BPT class$^b$  \\
& (J2000) & (J2000) & & (mag) &  & (\AA) & & (\AA) & & \\
(1) & (2) & (3) & (4) & (5)  & (7) & (8) & (9) & (10) & (15) & (16) \\
 \hline
587741490350587994 & 07:53:54.98 & +13:09:16.5 & 0.0476 & -22.33 & 0.54 & 5.20 & 0.27 & 5.00 & oETG & Quiescent\\
587739115235180592 & 08:02:15.80 & +18:24:08.3 & 0.0390 & -21.77 & 0.51 & 5.39 & 0.26 & 4.88 & pETG & Quiescent\\
587737809035526294 & 08:28:06.12 & +55:05:55.3 & 0.0392 & -21.80 & 1.28 & 4.75 & -0.30 & 1.24 & oLTG &  Star-forming\\
587742567860994239 & 10:42:22.37 & +18:08:06.8 & 0.0518 & -21.55 & 1.69 & 5.54 & -0.05 & 1.90 & pLTG & Liner\\
 \hline
 \end{tabular}
 \label{tab:all_sample}
(The full table is available in the online version of the paper. A portion of the table is shown here for guidance regarding form and content.)\\
 \end{center}
 \bf{~~Notes.}
\tablenotetext{a}{Observed line strength from OSSY catalogue.}
\tablenotetext{b}{Emission line classification.}
\end{table*}

Most importantly, assuming that our early-type NEOs are ``$\alpha$-enhanced'' , ``metal-rich'' 
and ``Na-enhanced''  without considering IMF variation, we found plausible parameters 
([$\alpha$/Fe]$\sim$\,0.3,  [Z/H]\,$\sim$\,0.3 and [Na/Fe]\,$\sim$\,0.3)  that reproduced the 
observed \mgb, \fe\ and \nad\ line strengths (see black filled star and red filled circle in 
Figure~\ref{fig:model}). A possible criticism is that the hypothesis of Na-enhancement to 
explain the Na excess is an \textit{ad hoc} assumption. However, several studies have pointed 
out the presence of Na-enhanced stars \citep[see e.g.][]{gsc04, fmr07}. We stress that 
Na-enhancement also causes an increase in Na\,{\small I} strength and counteracts the effect 
of the bottom-heavy IMF. 

Our analysis does not necessarily rule out the possibility of a bottom-heavy IMF. We simply 
report that \nad\ lines, though measured at a much higher S/N ratio than \na1, are too strong in 
many massive early-type galaxies to be accounted for by the same bottom-heavy IMF models 
that \citet{vc10} used to match the \na1\ lines. While Na enhancement is discussed here, 
we have not fully explored its effects on other line indices  (e.g. \mgb\ and \fe). This will be 
investigated in depth in a forthcoming paper based on new stellar population synthesis models. 
One more thing to pay attention to is the near-IR (NIR) flux. \citet{wis11} claimed that such 
bottom-heavy IMFs would yield redder NIR colors. In another forthcoming companion paper, 
we will also explore the multi-band (including NIR) photometric properties of NEOs.

ISM and dust could also increase \nad\ line strength. However, we classified the dust-lane 
galaxies as peculiar galaxies during the visual inspection stage (see Section~\ref{sec:visual}). 
Notwithstanding that diffuse dust is essentially invisible in optical imaging, our ordinary 
early-type NEOs are at least free from clumpy dust and emission lines (see 
Figure~\ref{fig:bpt_ETG} and Table~\ref{tab:bpt_ETG}). 
Furthermore, there was no correlation between \ebvs\  values from the OSSY catalogue 
and \fnad, at least for ordinary early-type NEOs, and the \ebvs\ values of ordinary early-type 
NEOs were nearly zero (see Figure~\ref{fig:ebv_ETG}). This implies that the ISM and/or dust 
does not enhance the \nad\ line strength of ordinary early-type NEOs in a significant way. 
Nevertheless, a study of the presence and properties of the ISM in early-type NEOs is 
desirable because it is known that a significant fraction of early-type galaxies contain some 
cool ISM and dust \citep[see e.g.][]{jetal87, ketal89}.

In contrast, the mechanism for \nad\ excess in late-type NEOs, including a small fraction of 
peculiar early-type NEOs that showed very similar trends to the late-type NEOs, appears to 
be different from that for the majority of early-type NEOs. In contrast to early-type NEOs, 
enhanced star formation was indicated both by strong H$\beta$ absorption line strengths 
and by the higher fraction that were classified as star-forming based on BPT diagnostics. An 
intriguing property worth discussing here is that late-type NEOs generally have weak \mgb\ 
and \fe\ line strengths. In that sense, these galaxies could correspond to objects like 
NGC\,3032 and 4150 observed in the course of the SAURON survey \citep[see e.g.][]{detal02} 
that experienced recent star formation \citep{jetal09}. These characteristics may be caused by 
young stars \citep{cetal11,ketal12}. The fact that late-type NEOs have extra young stars is not 
directly related to \nad\ line strength, because this would reduce \nad\ line strength rather than 
enhance it. However, the presence of star formation and nuclear activity in these galaxies 
implies the availability of gas and dust, which can impact \nad\ line strength. 
Furthermore, we found that our late-type NEOs tended to have larger \ebvs\ values than 
the late-type control sample (see Figure~\ref{fig:ebv_LTG}) in contrast to early-type NEOs.
  
We thus conclude that early-type (excluding a small fraction of peculiar early-type NEOs) and 
late-type NEOs have completely different mechanisms underlying their enhanced \nad\ 
strengths. The origin of \nad\ excess in early-type galaxies is not clear yet, but it is clear that 
early-type NEOs have Na-enhanced populations. The effects of Na-enhancement on 
individual line strengths will be elucidated in a companion paper. Meanwhile, \nad\ line 
strengths in late-type NEOs and a small fraction of peculiar early-type NEOs are highly 
contaminated by the ISM and/or dust. To facilitate follow-up observations of these exciting 
objects, we provide a catalogue of all \nad\ excess objects presented in this paper in 
Table~\ref{tab:all_sample}.

\section*{Acknowledgments}

The authors thank the anonymous referee for useful comments which led to 
improvements in the paper and acknowledge support from National Research 
Foundation of Korea (NRF-2009-0078756; NRF- 2010-0029391) and DRC Grant of 
Korea Research Council of Fundamental Science and Technology (FY 2012). Much 
of this manuscript was written during the visit of SKY to University of Nottingham and 
University of Oxford under the general support by LG Yon-Am Foundation. This project 
made use of the SDSS data.

\clearpage

\end{document}